\documentclass{JHEP3} 
\usepackage{amsmath}
\usepackage{epsfig}
\usepackage{amssymb,amsfonts}
\newcommand{\oao}[2]{{#1\atopwithdelims[]#2}}
\newcommand{\oaop}[2]{{#1\atopwithdelims()#2}}
\def\tf{\tilde{f}}
\def\tbf{\tilde{\bar{f}}}
\def\mq{\mathcal{Q}}
\def\ma{\mathcal{A}}

\def\mi{\mathcal{I}}
\def\mj{\mathcal{J}}
\def\mc{\mathcal{C}}
\def\mg{\mathcal{G}}
\def\mr{\mathcal{R}}
\def\mh{\mathcal{H}}

\def\tmt{\tilde{m}_t}
\def\tbmt{\tilde{\bar{m}}_t}

\def\zi{\mathbb{Z}}
\def\slr{SL(2,\mathbb{R})}

\title{
\boldmath 
The partition function of the supersymmetric two-dimensional 
black hole and little string theory
\thanks{Research partially supported by the EEC under the contracts
HPRN-CT-2000-00122, HPRN-CT-2000-00131.}
\unboldmath}
\author{Dan Isra\"el${}^1$, Costas Kounnas${}^1$, 
Ari Pakman${}^{1,2}$ and
Jan Troost${}^1$
\\  ${}^1$ Laboratoire de Physique Th\'eorique
de l'\'Ecole Normale Sup\'erieure\thanks{Unit{\'e} mixte  du
CNRS et de l'Ecole Normale Sup{\'e}rieure,
UMR 8549.}  \\ 24, Rue Lhomond  75231
Paris Cedex  05, France\\
and \\
 $ {}^2$ Racah Institute of Physics, The Hebrew University \\
Jerusalem 91904, Israel
$\ $ \\
E-mail:  \email{israel@lpt.ens.fr, kounnas@lpt.ens.fr, 
pakman@phys.huji.ac.il, troost@lpt.ens.fr }
\\
}

\abstract{
We compute the partition function of the supersymmetric two-dimensional
Euclidean black hole geometry 
described by the $SL(2,\mathbb{R})/U(1)$ 
superconformal field theory. 
We decompose the result in terms of characters 
of the $N=2$ superconformal symmetry. We point out puzzling sectors 
of states besides finding expected discrete and continuous contributions
to the partition function.
By adding an $N=2$ minimal model factor 
of the correct central charge and projecting on integral 
$N=2$ charges we compute the partition function of the background 
dual to little string theory in a double scaling limit. We show 
the precise correspondence between this theory 
and the background for NS5-branes on a circle, due to 
an exact description of the background as a null gauging of 
$\slr \times SU(2)$. Finally, we discuss the interplay 
between GSO projection and target space geometry.
}

\preprint{
LPTENS-04/16\\
RI-02-04
\\hep-th/0403237}

\begin{document}


\section{Introduction}
Conformal field theories with an $N=2$ superconformal 
symmetry have proven to be one of the most important
building blocks in superstring theory backgrounds. Firstly, a  
theorem of Banks and Dixon~\cite{Banks:1988yz} states 
than any superstring vacua with (at least) $N=1$ spacetime 
supersymmetry in four dimensions is built by combining 
internal conformal field theories with worldsheet 
$N=2$ superconformal symmetry.\footnote{See~\cite{Giveon:2003ku} for 
a generalization to $AdS_3 \times \mathcal{M}$ backgrounds.} A large class of
examples is 
provided by Gepner models~\cite{Gepner:1987qi}, 
that use the minimal $N=2$ theories -- constructed 
with the $SU(2) / U(1)$ coset -- to obtain non-trivial 
supersymmetric string compactifications.
The $N=2$ algebra was also used to study Calabi-Yau 
compactifications in general, leading in particular to the 
further development
 of mirror symmetry~\cite{Dixon:1987bg}~\cite{Lerche:1989uy}.
 
The non-minimal $N=2$ theories have been shown 
in~\cite{Dixon:1989cg} to be intimately related 
to the (bosonic) coset theory $\slr / U(1)$. In this 
paper we will elaborate on this connection by 
identifying these non-minimal 
$N=2$ representations with the coset representations of the 
supersymmetric $\slr / U(1)$ coset~\cite{Nojiri:1991hb}, 
i.e. the gauged supersymmetric WZW model. The axial coset 
has a two-dimensional target space with a cigar 
geometry, and in two dimensions it has an interpretation as
a Euclidean black hole~\cite{Witten:1991yr}. Recent interest for this 
superconformal field theory derives from the 
duality with the $N=2$ super-Liouville theory, 
conjectured in~\cite{Giveon:1999px} and 
studied in~\cite{Hori:2001ax}~\cite{Tong:2003ik}. 
The present work casts some 
new light on this duality, because we show 
explicitely that the partition function  
of the supersymmetric $\slr / U(1)$ can be 
decomposed in terms of (extended) $N=2$ 
characters, which are the buildings blocks of 
the $N=2$ super-Liouville theory~\cite{Eguchi:2003ik}. 
This is natural in the light of the fact that the characters
for the two models are identical.
Another motivation for this work comes from 
the double scaling limit of the little string 
theory~\cite{Giveon:1999px}, corresponding to 
NS5-branes spread on a (topologically trivial) circle in the transverse space. 
In~\cite{Giveon:1999px} the worldsheet CFT corresponding 
to the bulk geometry was argued to be the orbifold
$$\frac{\slr_{k} / U(1) \times SU(2)_{k} / U(1)}{\zi_{k}}.$$ 
In this work we compute the partition function for this theory, 
and show explicitly that the $\zi_{k}$ is completely fixed by the 
requirement of spacetime supersymmetry, without any 
a priori assumption
on the bulk geometry. Then we argue that this projection 
changes drastically the low-energy interpretation of the background, 
and that this orbifold is actually equivalent to a null 
gauged WZW model:
$$\frac{\slr_{k} \times SU(2)_{k}}{U(1)_L \times U(1)_R},$$
whose target space corresponds precisely to a configuration of 
five-branes on a circle in supergravity, previously written down 
in~\cite{Sfetsos:1998xd}. 
\subsection*{Note added}
When this work was near completion, the interesting work~\cite{ES} 
appeared on the archive with a large overlap with the present 
paper. Among the complementary issues we consider are: 
the supersymmetric marginal deformation method, 
a precise analysis of the partition function
and its decomposition in characters (including multiplicities of non-primary
states)
and the relationship with the background for five-branes on a circle.
On the other hand \cite{ES} has other additional and interesting
material especially 
regarding the elliptic genus for these non-compact models.

\section{The supersymmetric coset by deformation} 
Our aim is to find the modular invariant torus partition 
function for the supersymmetric $\slr / U(1)$ coset conformal 
field theory. 
The strategy we will use is to deform the supersymmetric
$SL(2,\mathbb{R})_k$ WZW model by a marginal deformation that 
preserves supersymmetry.
It is a very economic way of obtaining the 
supersymmetric partition function for the supersymmetric coset. As 
a byproduct, it is interesting on its own right since supersymmetric 
deformations of $\slr$ have only been
studied in the asymmetric case (see~\cite{Israel:2003cx}). 
In the bosonic case, the partition function of the coset 
has been computed in~\cite{Hanany:2002ev}. 
The discrete part has been also studied in~\cite{Hwang:1992uk}.
  
\boldmath
\subsection{Partition function for supersymmetric $\slr$}
\unboldmath
The partition function for the $N=1$ supersymmetric $\slr$  
at level $k$ is made of the direct
product of a modular invariant factor for the purely bosonic 
$\slr$ at level $k+2$ and a modular invariant\footnote{Or modular covariant 
factor for superstrings.} factor for the free fermions, realizing a 
level -2 $\slr$ algebra. The purely bosonic 
partition function has been obtained in~\cite{Israel:2003ry}
(following methods in \cite{Maldacena:2000kv}\cite{Hanany:2002ev}). We 
refer the reader to these works for more detail. 
Let us start with the type $0B$ partition function (i.e. the partition
function
diagonal in fermion boundary conditions), for the universal 
cover of the $\slr$ algebra -- i.e. $AdS_3$:
\footnote{We have removed the contribution of the 
neutral fermions associated to left and right $U(1)$ currents
$J^3$ and $\bar{J}^3$ since they play 
no role in the following derivation.} 
\begin{eqnarray}
Z & = &  \frac{\sqrt{k\tau_2}}{\eta \bar{\eta}} 
\int d^2 s\  \frac{e^{2\pi \tau_2 s_{1}^2}}{\left|
\vartheta_1 (s_1 \tau - s_2 | \tau ) \right|^2} \nonumber \\
&&\times \ \int_0^1 dt_1 \ dt_2 \  \sum_{n,w \in \mathbb{Z}}
e^{2i\pi n s_2}
q^{\frac{1}{4(k+2)} \left( n- (k+2)(w+s_1-t_1) \right)^2}
\bar{q}^{\frac{1}{4(k+2)} \left( n+ (k+2)(w+s_1-t_1) \right)^2} \nonumber \\
&&\times \ \sum_{\tilde{n},\tilde{w}} 
e^{2i\pi (n-\tilde{n})t_2}
q^{-\frac{1}{4(k+2)} \left( \tilde{n}+ (k+2)(\tilde{w} + t_1 ) \right)^2}
\bar{q}^{-\frac{1}{4(k+2)} \left( \tilde{n}- (k+2)(\tilde{w} + t_1 )\right)^2}
\nonumber \\
&&\times \ \frac{1}{2} \sum_{a,b=0}^{1} \vartheta \oao{a}{b} (\tau,0) \ \bar{\vartheta}
\oao{a}{b} (\tau,0)\, .
\label{partsl2}
\end{eqnarray}
The first two lines are the partition function for the 
bosonic coset $\slr / U(1)$ while the third line corresponds 
to the extra time-like $U(1)$ that extends the coset theory 
to $\slr$. They are coupled through the $\mathbb{Z}$ orbifold 
given by the integral over $t_1$ and $t_2$.\footnote{At least 
for integer level $k$, the partition function for the single cover 
is obtained by using instead a $\zi_{k+2 }$ orbifold. But, as it is explained 
in~\cite{Israel:2003ry}, any cover of $\slr$ gives the same axial coset.}
The fermionic
characters are defined as usual:
$$\vartheta \oao{a}{b} (\tau ,\nu  ) = \sum_{f \in \zi} 
q^{\frac{1}{2}\left(f+\frac{a}{2} \right)^2} e^{2i\pi (\nu + b/2)(f+a/2)}.$$
To proceed, we first integrate over $t_2$, and obtain
the constraint $n=\tilde{n}$. Then we introduce the spectral flow 
quantum number: $w_+ = w + \tilde{w}$. 
Finally we factorize the supersymmetric $\mj^3 \bar \mj^3$ lattice 
associated to the total currents
at level $k$, that will eventually be removed by the gauging:
\begin{eqnarray}
Z & = &  \frac{\sqrt{k\tau_2}}{\eta \bar{\eta}} 
\int_{0}^{1} ds_1 \ ds_2 \  \frac{e^{2\pi \tau_2 s_{1}^2}}{\left|
\vartheta_1 (s_1 \tau - s_2 | \tau )^2 \right|^2} 
\nonumber \\
&&\times  \ \int_{0}^{1} dt_1 \  \sum_{n,w \in \mathbb{Z}}
e^{2i\pi n s_2}\nonumber \\
&&\times \ q^{\frac{1}{4(k+2)} \left( n- (k+2)(w+s_1-t_1) \right)^2}
\bar{q}^{\frac{1}{4(k+2)} \left( n+ (k+2)(w+s_1-t_1) \right)^2}
\nonumber \\
&&\times \ \sum_{f,\bar{f},w_+ \in \mathbb{Z}} 
\frac{1}{2} \sum_{a,b \in \mathbb{Z}_2} e^{i\pi (f-\bar{f}) b } 
\nonumber \\  
&&\times \ q^{\frac{k}{2(k+2)} \left( f+ \frac{a}{2} + \frac{
n-(k+2)(w-t_1)+(k+2)w_+ +2f +a}{k} \right)^2}
q^{-\frac{1}{k} \left( 
\frac{n}{2} - \frac{k+2}{2}(w-t_1) + \frac{k+2}{2}w_+ 
+f +\frac{a}{2} \right)^2} \nonumber \\
&&\times \ \bar{q}^{\frac{k}{2(k+2)} \left(\bar{f} + \frac{a}{2} + \frac{
-n-(k+2)(w-t_1)+(k+2)w_+ +2\bar{f} +a}{k} \right
)^2}
\bar{q}^{-\frac{1}{k} \left( 
-\frac{n}{2} - \frac{k+2}{2}(w-t_1) + \frac{k+2}{2}w_+ 
+\bar{f} +\frac{a}{2} \right)^2}
\end{eqnarray}
The total current of the elliptic sub-algebra is given by the 
sum of the purely bosonic current and the fermionic current: 
$\mathcal{J}^3 = J^3 + : \psi^+ \psi^- :$, where the fermions
have charge $\pm 1$ under the global $U(1)$. 
\subsection{Deformation towards the supersymmetric coset}
A marginal deformation compatible with the (local) 
$N=1$ superconformal symmetry of the supersymmetric 
$SL(2,\mathbb{R})$ theory is given 
by the current-current operator constructed with the 
{\it total} currents of the Cartan sub-algebra.\footnote{
We have chosen as a Cartan the elliptic sub-algebra, 
in order to obtain at the end the Euclidean black hole. 
For the Minkowskian case one has to consider instead the hyperbolic 
(non-compact) one.} The action is perturbed by the operator:
$$\delta S \sim \int d^2 z \ \mj^3 (z) \bar \mj^3 (\bar{z} ).$$
The action of this deformation at the level of the partition function 
is obtained by changing the radius of the corresponding lattice of  
$(\mj^3 ,\bar \mj^3)$, which we call $(P_L , P_R )$ in the 
following. We replace:
\begin{eqnarray*}
\frac{1}{2} P_{L}^2 
& \to & - \frac{1}{4k} 
\left[ \left(n+f-\bar{f} \right) \mathcal{R} 
\ + \ \left( -(k+2)(w -t_1)  + (k+2)w_+ + f + \bar{f} + a 
\right) / \mathcal{R} 
\right]^2
\\
\frac{1}{2} P_{R}^2 
& \to & - \frac{1}{4k} 
\left[ - \left(n+f -\bar{f} \right)  \mathcal{R} 
\ + \ \left( -(k+2)(w -t_1)  + (k+2)w_+ + f + \bar{f} +a
\right) / \mathcal{R} 
\right]^2 .
\end{eqnarray*}
The modular invariance is ensured along the line of deformation 
since it is a $O(1,1,\mathbb{R})$ transformation of a toroidal 
lattice. This should be understood in a formal sense 
since it is a divergent sum over timelike momenta. 
To go from the parent theory $\slr$ to the coset theory 
$\slr / U(1)$ we have to remove all the contributions 
of the $U(1)$ which is gauged. It is known, at least for 
the bosonic case~\cite{Giveon:1993ph}~\cite{Israel:2003ry}, 
that it can be done in a modular invariant way by taking 
an {\it infinite} marginal deformation, either 
$\mr \to 0$ (axial coset) or $\mr \to \infty$ (vector coset). 
In those limits the contribution of the extra $U(1)$ decouples 
and can be safely removed. Here, since we would like to find the 
partition function for the cigar - the axial coset, we have to 
take the limit where the radius goes to zero. 
We have to {\it (i)} do an analytic continuation to get a well-defined 
sum over the momenta : $\mr \to i \mr$ and {\it (ii)} take the limit 
$\mr \to 0$. The validity of this procedure will be 
checked later by showing explicitely the modular invariance 
of the resulting partition function.  
In the zero radius limit (i.e. axial coset), we get 
the constraint: 
\begin{equation}
\label{constr}
-(k+2)(w-t_1)+(k+2)w_+ + f+\bar{f}+a =0
\end{equation}
We define also $N=n+f - \bar{f} $. 
Then we get the following partition function, after 
dropping the decoupled contribution of the extra $U(1)$: 
\begin{eqnarray}
Z & = &  \sqrt{k \tau_2}\ 
\frac{1}{2} \sum_{a,b \in \mathbb{Z}_2}
\int_{0}^{1} ds_1 \ ds_2 \  \frac{e^{2\pi \tau_2 s_{1}^2}}{\left|
\vartheta_1 (s_1 \tau - s_2 | \tau )^2 \right|^2} \nonumber \\
&& \times \ 
\sum_{f,\bar{f},N,w_+ \in \mathbb{Z}} 
e^{i\pi (f-\bar{f} ) b } 
e^{2i\pi (N-f +\bar{f}) s_2}
\nonumber \\
&& \times \ q^{\frac{1}{4(k+2)} \left( N-2f -a - (k+2)(w_+ + s_1) 
\right)^2}
\bar{q}^{\frac{1}{4(k+2)} \left( N+2\bar{f} + a + (k+2)(w_+ + s_1) 
\right)^2} \nonumber \\
&&\times \ q^{\frac{k}{2(k+2)} \left( f+ \frac{a}{2} + \frac{N}{k} \right)^2}
\bar{q}^{\frac{k}{2(k+2)} \left(\bar{f} + \frac{a}{2} - \frac{N}{k} \right
)^2}.
\label{partco}
\end{eqnarray}
This is the partition function for the supersymmetric coset 
$\slr_{k} / U(1)$ we wanted to compute. The central charge 
of this superconformal field theory is~: 
$$c = \frac{3(k+2)}{k}.$$
We recall that the N=2 R-current of the super-coset, obtained 
by the standard Kazama-Suzuki procedure~\cite{Kazama:1988qp}
is: 
$$J_{R} = :\psi^+ \psi^- : + \frac{2 \mathcal{J}^3}{k} 
$$
It is orthogonal to the total current $\mathcal{J}^3$. 
The last line of the partition function, eq.~(\ref{partco}) 
corresponds to the lattice 
of this $N=2$ R-current by construction. So the left and right N=2 charges 
of the super-coset are: 
\begin{equation}
\mq_L =  f+ \frac{a}{2} + \frac{N}{k} \ , \ \ 
\mq_R =  \bar{f} + \frac{a}{2} - \frac{N}{k}
\end{equation}
\subsection{Modular invariance}
Our result is by construction modular invariant since it has 
been obtained as the limiting case of a line of modular 
invariant theories. Nevertheless we will check the modular 
invariance directly, by putting the partition function 
in a form that will be at any rate useful in the following. 
\subsubsection*{Partition function in Lagrangian form}
First we factorize from the R-current lattice the contribution 
that came from the free fermions in the parent theory:
\begin{eqnarray}
Z  =   \sqrt{k\tau_2}
\frac{1}{2} \sum_{a,b \in \mathbb{Z}_2}
\int_{0}^{1} ds_1 \ ds_2 \   
\sum_{f,\bar{f},N,w_+ \in \mathbb{Z}} 
e^{i\pi (f-\bar{f} ) b } \nonumber \\
\times \ e^{2i\pi (N-f +\bar{f})  s_2} 
e^{-2i\pi \tau_1 (N-f+\bar{f} ) s_1} 
e^{-2\pi \tau_2 (f+\bar{f} + a+(k+2)w_+ )s_1}
e^{-\pi \tau_2 k s_{1}^2}
\nonumber \\
\times \ \frac{1}{\left|
\vartheta_1 (s_1 \tau - s_2 | \tau )^2 \right|^2}\ 
q^{ \frac{1}{2} \left( f +\frac{a}{2} + w_+ 
\right)^2 + \frac{(N-kw_+)^2}{4k} }
\bar{q}^{  \frac{1}{2} \left( \bar{f} + 
\frac{a}{2} + w_+ 
\right)^2 + \frac{(N+kw_+)^2}{4k}}\nonumber \\
\label{GKO}
\end{eqnarray}
Then we introduce the flowed fermionic levels, since the spectral 
flow also has to act on the $\slr$ algebra at level $-2$ made with  
the fermions, as explained in~\cite{Pakman:2003cu}:
$$f+w_+ =\tf  \ , \ \ \bar{f} +w_+ = \tbf .$$
Recast in this form, namely eq.~(\ref{GKO}), the structure of the partition 
function is transparent. 
In fact, the weights that will appear in the spectrum will have the following 
form:
$$L_0 = L_{0}^{SL(2,\mathbb{R}),bosonic} 
+ L_{0}^{\slr, fermionic} - L_{0}^{U(1)_{-k}},$$
which is exactly what is expected from the standard GKO 
construction~\cite{Goddard:ee}.\\
To prove modular invariance, we observe that the partition function, 
(\ref{GKO}) can be rewritten in the Lagrangian form, 
by a Poisson resummation on the momenta $N$ of the compact boson 
as follows:
\begin{eqnarray}
Z =  \frac{k}{2} \sum_{a,b \in \mathbb{Z}_2}
\int_{0}^{1} ds_1 \ ds_2 \
\frac{e^{2\pi \tau_2 s_{1}^2}}{\left|
\vartheta_1 (s_1 \tau - s_2 | \tau )^2 \right|^2} 
\sum_{M,w_{+} \in \mathbb{Z}}
e^{-\frac{k\pi}{\tau_2} \left|(w_+ + s_1 )\tau 
-(M+s_2)\right|^2} \nonumber\\
\times \ \vartheta \oao{a+2s_1}{b+2s_2} 
\bar{\vartheta} \oao{a+2s_1}{b+2s_2} . \nonumber\\
\label{lagr}
\end{eqnarray}
The modular invariance of this expression is straightforwardly 
checked since it has the structure of a freely acting 
(continuous) orbifold (see~\cite{Israel:2003ry} for more 
details). The partition function~(\ref{lagr}) 
is clearly understood as the constrained product 
of $SL(2,\mathbb{R})_{k+2} \times U(1)_{k} \times U(1)_2$.

\subsubsection*{The twisted super-coset partition function}
The super-coset $\slr / U(1)$ has a discrete 
chiral symmetry $\zi_2 \times \zi$.
While the first factor corresponds to the usual $\zi_2$ symmetry of the fermions, 
the second factor is the symmetry of the non-compact supersymmetric 
parafermions. Any discrete subgroup of this $\zi$ symmetry can
be gauged. We will need for the following the $\mathbb{Z}_{k}$ orbifold 
of this theory. Geometrically, 
this orbifold acts on the translational symmetry along 
the transverse compact coordinate of the cigar -- the Euclidean 
compact time direction of the 2D black hole. 
It is obtained by acting in the lattice of 
$U(1)_{k}$ as follows: 
\begin{eqnarray}
Z_{orb} &=& \frac{1}{k} \sum_{\gamma, \delta \in  \mathbb{Z}_{k}}
Z_{BH} \oao{\gamma}{\delta}
\nonumber \\ 
&=&  \sum_{\gamma,\delta \in  \mathbb{Z}_{k}}
\frac{1}{2} \sum_{a,b \in \mathbb{Z}_2}
\int_{0}^{1} ds_1 \ ds_2 \
\frac{e^{2\pi \tau_2 s_{1}^2}}{\left|
\vartheta_1 (s_1 \tau - s_2 | \tau )^2 \right|^2} 
\nonumber \\ 
&& \times \ \sum_{M,w_{+} \in \mathbb{Z}}
e^{-\frac{k\pi}{\tau_2} \left|
\left( w_+ + s_1 + \frac{\gamma}{k} \right)\tau 
-\left( M+ s_2 + \frac{\delta}{k}\right) \right|^2} 
\vartheta \oao{a+2s_1}{b+2s_2} 
\bar{\vartheta} \oao{a+2s_1}{b+2s_2}
\end{eqnarray}
The modular invariance of this expression is manifest. 
In the Hamiltonian representation (i.e. after Poisson resummation), 
we will have in eq. \ref{partco}: 
{\it (i)} to do the replacement $w_+ \to w_+ + \gamma/k$ 
and {\it (ii)} to insert the phase  
$e^{2i\pi N \delta/k}$. So we obtain the following twisted sectors:
\begin{eqnarray}
Z_{BH} \oao{a;a}{b;b}\oao{\gamma}{\delta} & = &  \sqrt{k \tau_2}
\int_{0}^{1} ds_1 \ ds_2 \  \frac{e^{2\pi \tau_2 s_{1}^2}}{\left|
\vartheta_1 (s_1 \tau - s_2 | \tau )^2 \right|^2} \nonumber \\
&& \times \ 
\sum_{\tf,\tbf,N,w_+ \in \mathbb{Z}}
e^{i\pi (\tf-\tbf ) b } 
e^{2i\pi (N-\tf +\tbf) s_2}
\nonumber \\
&& \times \ q^{\frac{1}{4(k+2)} \left( N-2\tf -a - kw_+ - \gamma  -(k+2)s_1 
\right)^2}
\bar{q}^{\frac{1}{4(k+2)} \left( N+2\tbf + a + kw_+ + \gamma +(k+2)s_1  
\right)^2} \nonumber \\
&& \times \ q^{\frac{k}{2(k+2)} \left( \tf + \frac{a}{2} + 
\frac{N-\gamma}{k} -w_+\right)^2}
\bar{q}^{\frac{k}{2(k+2)} \left(\tbf + \frac{a}{2} -
  \frac{N+\gamma}{k} -w_+ 
\right)^2} e^{2i\pi N \frac{\delta}{k}}
\label{partcotw}
\end{eqnarray}
In this expression $\oao{a;\bar{a}}{b;\bar{b}}$ denotes the 
spin structures on the two-torus for the left and right-movers. 
If the left and right spin structures are not the same, one 
has to insert an additional phase $e^{-i\pi (b - \bar{b} )s_1}$
 to ensure modular 
invariance of the orbifold.


\section{The character decomposition}
\label{chardecsect}
The analysis of the partition function we obtained for the 
super-coset, eq.~(\ref{partco}) is carried out by identifying the 
characters of the super-coset $\slr / U(1)$ corresponding to the different 
representations, and then the irreducible characters of the 
(non-minimal) $N=2$ superconformal algebra. In the following analysis,
we will improve on the analyses that are present in the literature
in the fact that we will take along multiplicities of all the states
in the partition function explicitly.


\subsection{The discrete representations}
The first step in this decomposition is to expand the 
$\vartheta_1$ factor coming from the twisted determinant 
of Euclidean $AdS_3 = SL(2,\mathbb{C}) / SU(2)$ 
in powers of the holonomies of the gauge field. Explicitely, we have, with
$y = e^{2i\pi (s_1 \tau - s_2)}$:
\begin{equation}
\frac{1}{\vartheta_1 (s_1 \tau - s_2 |\tau)} 
=  \frac{i\ y^{1/2} q^{-1/12}}{\eta (\tau )} 
\frac{1}{\prod_{n=1}^{\infty} (1-yq^{n-1})(1-y^{-1} q^n )}
=  \frac{i\ y^{1/2}}{\eta^3} \sum_{r \in \zi} y^r S_r (\tau )
\label{decomp}
\end{equation}
where the function $S_r$ is defined by (see~\cite{Pakman:2003kh}):
\begin{equation}
S_r (\tau ) = \sum_{n=0}^{\infty} (-)^n q^{\frac{n(n+2r+1)}{2}}
\label{srfct}
\end{equation}
Note that the above expansion, eq.~(\ref{decomp}) is valid only 
for $|q|<|y|<1$ which is indeed the case when $1>s_1 >0$. We plug this expansion in the partition 
function, eq.~(\ref{partco}), and we also {\it(i)} integrate over 
$s_2$ to get the constraint: $N=r-\bar r+\tf - \tbf$ and {\it (ii)} introduce 
a Gaussian integral over $s$ to linearize the integral over 
$s_1$, as was done in~\cite{Hanany:2002ev} for the bosonic coset. 
Then we obtain the following expression: 
\begin{eqnarray}
Z & = &  (q\bar{q})^{-\frac{c-3}{24}} 
\frac{\tau_2}{\eta^3 \bar{\eta}^3}
\sum_{a,b \in \mathbb{Z}_2}
\int_{-\infty}^{\infty} ds\  
\int_{0}^{1} ds_1 \   
\sum_{\tf,\tbf,N,w_+ \in \mathbb{Z}} 
e^{i\pi (\tf-\tbf) b } \nonumber \\
&& \times \ \sum_{r,\bar{r} \in \mathbb{Z}}\ 
e^{-2\pi \tau_2 (2is+1+r+\bar{r}+\tf+\tbf+ a+kw_+ )s_1}
\delta_{r-\bar{r}+\tf-\tbf,N} \ 
S_r  \bar{S}_{\bar{r}} 
\nonumber \\
&&\times \ q^{\frac{s^2 + 1/4}{k} + 
\frac{1}{4(k+2)} \left( -N+2\tf +a + kw_+ \right)^2 
+ \frac{k}{2(k+2)} \left( \tf + \frac{a}{2} + \frac{N-kw_+}{k} \right)^2}
\nonumber \\
&&\times \ \bar{q}^{\frac{s^2 + 1/4}{k} + \frac{1}{4(k+2)} 
\left( N+2\tbf + a + kw_+  \right)^2 
+ \frac{k}{2(k+2)} \left(\tbf + \frac{a}{2} - \frac{N+kw}{k} \right
)^2}\nonumber \\
\label{partlin}
\end{eqnarray}
For the discrete representations first, we can proceed to the character 
decomposition. Starting from~(\ref{partlin}), we first integrate over $s_1$ 
to obtain: 
\begin{eqnarray}
Z & = & \frac{1}{\pi} (q\bar{q})^{-\frac{c-3}{24}} 
\frac{1}{\eta^3 \bar{\eta}^3}
\frac{1}{2} \sum_{a,b \in \mathbb{Z}_2}
\int_{-\infty}^{\infty} ds\  
\sum_{\tf,\tbf,N,w_+ \in \mathbb{Z}} 
e^{i\pi (\tf-\tbf) b } \nonumber \\
&& \times \ \sum_{r,\bar{r} \in \mathbb{Z}}\ 
\delta_{r-\bar{r}+\tf-\tbf,N} \ 
S_r  \bar{S}_{\bar{r}} 
\nonumber \\
&& \times \ q^{\frac{s^2 + 1/4}{k} + 
\frac{1}{4(k+2)} \left( -N+2\tf +a + kw_+ \right)^2 
+ \frac{k}{2(k+2)} \left( \tf + \frac{a}{2} + \frac{N-kw_+}{k} \right)^2}
\nonumber \\
&& \times \ \bar{q}^{\frac{s^2 + 1/4}{k} + \frac{1}{4(k+2)} 
\left( N+2\tbf + a + kw_+  \right)^2 
+ \frac{k}{2(k+2)} \left(\tbf + \frac{a}{2} - \frac{N+kw}{k} \right
)^2}\nonumber \\
&& \times \ \frac{1-e^{-2\pi \tau_2 (2is+1+r+\bar{r}+\tf+\tbf+ a+kw_+ )}}{
2is+1+r+\bar{r}+\tf+\tbf+ a+kw_+ }
\label{partdecd}
\end{eqnarray}

\subsubsection*{Discrete representations of the coset}
The {\it discrete representations} of $\slr$ are either 
highest weight or lowest weight. When we parameterize the 
Casimir as: $c_2 = s^2 + 1/4$, they are labeled by 
$s=i(1/2 -j)$, where $j \in \mathbb{R}_+$ for the unitary 
representations. The spectrum of the elliptic sub-algebra 
(i.e. $J^3$) is: $m=\pm (j+r)$, $r \in \mathbb{N}$.
We have also to consider {\it flowed representations} 
(see~\cite{Henningson:1991jc}\cite{Maldacena:2000hw}), for which the quantum numbers
$m$ equal $m = \tilde{m} + 
kw_+/2$. 

To find the contribution of the discrete representations to 
the partition function, we observe that the second term in the last 
line of eq.~(\ref{partdecd}) can be completed to 
a square if we shift the variable $s$ as: $s \to  s - ik/2$. 
We pick up residues between the two contours for: 
$$2\Im s = 1+ r+\bar{r}+ \tf+\tbf+ a+kw_+ .$$ 
These poles will correspond to the discrete representations of the coset. 
We find that they are located in the interval: 
\begin{equation}
\frac{1}{2} <j< \frac{k+1}{2},
\label{rangej}
\end{equation}
which is precisely the improved unitarity range (for level $k+2$) 
for the discrete representations found in~\cite{Maldacena:2000hw}.

\boldmath
\subsubsection*{Representations $j=1/2$ and $j=(k+1)/2$.}
\unboldmath
The expression obtained after the integration over $s_1$, 
eq.~(\ref{partdecd}), has a pole for $s=0$ whenever we have:
$$1+r+\bar{r}+\tf+\tbf+ a+kw_+ =0.$$ 
This pole corresponds to the discrete representation with 
$j=1/2$. Thus we choose to define the integral over 
$s$ with the principal value prescription 
(written as usual $P\int$), taking into 
account half of the pole. After the contour shift in the second 
term in the last line of eq.~(\ref{partdecd}), we will find also 
half of the pole corresponding to the representation $j=(k+1)/2$.
\subsubsection*{Discrete characters of the bosonic coset}
To summarize, we can write the contribution of the discrete spectrum as: 
\begin{eqnarray}
Z_{D} &=& \frac{1}{\eta \bar{\eta}} \ 
\frac{1}{2}\sum_{a,b \in \mathbb{Z}_2}\ 
\int_{1/2}^{(k+1)/2} dj \ 
\sum_{r,\bar{r},\tf,\tbf,N,w_+ \in \mathbb{Z}} 
e^{i\pi (\tf-\tbf ) b}\ 
\lambda^{j}_{r} (q) \ \lambda^{j}_{\bar{r}} (\bar{q})\nonumber \\ 
&&\times \ q^{\frac{k}{2(k+2)} \left( 
\tf + \frac{a}{2} + \frac{N-kw_+}{k} \right)^2}
\bar{q}^{\frac{k}{2(k+2)} \left( 
\tbf + \frac{a}{2} + \frac{-N-kw_+}{k} \right)^2}
\nonumber \\
&& \times \ \delta (2j+r+\bar{r}+kw_+ +\tf +\tbf +a ) \ 
\delta_{r-\bar{r}+\tf - \tbf ,N}
\end{eqnarray}
where the discrete characters of the bosonic coset $\slr / U(1)$ 
are defined as (for several derivations of these characters, see the 
references~\cite{Sfetsos:1991wn}\cite{Bakas:1991fs}\cite{Griffin:1990fg}\cite{Pakman:2003kh}\cite{Ribault:2003ss}):
\begin{equation}
\lambda^{j}_{r} (q) = q^{-\frac{(j-1/2)^2}{k} + \frac{(j+r)^2}{k+2}}
\frac{S_r (\tau )}{\eta^2 (\tau )}.
\end{equation}
Note that we have found not only the correct conformal weights for the primaries, 
but also the right multiplicities for all the descendents contained in the characters. 
The discrete part of the partition function is thus composed of the product of 
discrete characters of the bosonic coset $\slr / U(1)$ 
and $U(1)$ characters of the $N=2$ 
R-current, obeying some constraints. The constraints on the $N=2$ 
charges are those expected for 
the path integral construction of the supersymmetric 
coset~\cite{Schnitzer:1988qj}, namely ($m_t$ is the eigenvalue of $\mj^3$):
\begin{equation}
m_{t} + \bar{m}_{t} = \tilde{m}_t + \tilde{\bar{m}}_t 
+ kw_+ 
= 2j+r+\bar{r}+kw_+ +\tf +\tbf + a = 0,
\end{equation}
and 
\begin{equation}
m_{t} - \bar{m}_{t} 
= r-\bar{r}+ \tf - \tbf  = N,
\end{equation}
The solution of these constraints is: 
$m_{t} = N/2 \ \  \mathrm{and} \ \bar{m}_{t} = - N/2$, or, for 
the unflowed quantum numbers: 
\begin{equation}
\tilde m_{t} = \frac{N}{2} -  \frac{k}{2}w_+ \ \  \mathrm{and} \ \ 
\tilde{\bar{m}}_{t} = - \frac{N}{2} -\frac{k}{2}w_+.
\label{paramq}
\end{equation}
To proceed we can solve explicitely for the constraints: 
\begin{itemize}
\item first we trade the integral over $j$ by a sum over half-integer 
values within the range~(\ref{rangej})
\item then we solve the constraints for $r$ and $\bar{r}$ for 
a given $j$: 
\begin{eqnarray}
r &=& \frac{N}{2} - \frac{k}{2}w_+ - \left( \tf + \frac{a}{2} 
\right) -j \nonumber\\
\bar{r} & = & -\frac{N}{2} - \frac{k}{2}w_+ - \left( \tbf 
+ \frac{a}{2} \right)  -j
\label{solvcont}
\end{eqnarray}
\end{itemize} 
Since $r$ and $\bar{r}$ have to be integers, we will define in the 
following the discrete coset characters to be zero if the 
solutions of the constraints, eq.~(\ref{solvcont}) are not 
integer-valued. 
With this convention in mind, we can recast the contribution of the 
discrete representations to the partition function as follows:
\begin{eqnarray}
Z_{D} &=& \frac{1}{\eta \bar{\eta}} (q\bar{q})^{-c/24} 
\frac{1}{2} \sum_{a,b \in \mathbb{Z}_2}
\sum_{2j=1}^{k+1} \Upsilon ( 2j,1 )
\sum_{N,w_+ \in \mathbb{Z}} \nonumber \\ 
&& \times \ \sum_{\tf \in \mathbb{Z}} e^{i\pi b \tf}\ 
\lambda^{j}_{\frac{N}{2} - \frac{k}{2}w_+ -
\left(\tf + \frac{a}{2} \right) -j}   
 \  q^{\frac{k}{2(k+2)} \left( 
\tf + \frac{a}{2} + \frac{N-kw_+}{k} \right)^2}
\nonumber \\ 
&&\times \ \sum_{\tbf \in \mathbb{Z}}  e^{-i\pi b \tbf}\ 
\bar{\lambda}^{j}_{-\frac{N}{2} - \frac{k}{2}w_+ -\left( \tbf 
+ \frac{a}{2} \right) -j}  \  
\bar{q}^{\frac{k}{2(k+2)} \left( 
\tbf + \frac{a}{2} + \frac{-N-kw_+}{k} \right)^2} \, .
\label{partdisc}
\end{eqnarray}
In this expression we have defined:
\begin{eqnarray}
\Upsilon (2j,1 ) &=& \frac{1}{2} \ \ \mathrm{if} \ 2j=1 \mod k 
\nonumber \\
& =& 1 \quad \mathrm{otherwise.} 
\end{eqnarray}

\subsubsection*{Link with extended N=2 discrete characters}
In this section we would like to make the connection to the decomposition
of $N=2$ extended characters in term of the characters of the bosonic
coset ~\cite{Israel:2004xj}. In fact, from this analysis it is clear
that the extended $N=2$ characters and the extended characters for the
supercoset agree. First, we observe that the partition function of the 
discrete representations, eq.~(\ref{partdisc}), is written 
in terms of the following (twisted) unextended $N=2$ discrete characters: 
\begin{eqnarray}
ch_d (j,r) \oao{a}{b} (\tau,\nu ) &=& 
\frac{1}{\eta^3 (\tau )} 
q^{- \frac{(j-1/2)^2+(j+r+a/2)^2}{k}} z^{\frac{2j+2r+a}{k}}
\frac{\vartheta \oao{a}{b} (\tau, \nu )}{1+(-)^b z\ q^{(1+a)/2+r}}\nonumber\\ 
&=& \sum_{n \in \mathbb{Z}} e^{i\pi b \left(n + \frac{a}{2} \right)}
\lambda^{j}_{\frac{k}{2} Q_m - n - \frac{a}{2} -j}  
 \  \frac{q^{\frac{k}{2(k+2)} \left( 
n + \frac{a}{2} + Q_m \right)^2}}{\eta (\tau )} z^{n + \frac{a}{2} + Q_m}
\label{unext2}
\end{eqnarray}
where $$Q_m = \frac{2m}{k} = \frac{2(j+r)}{k}$$ 
is the $N=2$ charge of the primary, and $z= e^{2i\pi\nu}$. 
Then we rewrite the discrete partition function as: 
\begin{eqnarray}
Z_{D} =  \frac{1}{2} 
\sum_{a,b \in \mathbb{Z}_2}\ 
\sum_{2j=1}^{k+1} \ \Upsilon (2j,1  )\  
\sum_{N,w_+ \in \mathbb{Z}} \nonumber \\ 
\times \ ch_d \left(j,\frac{N}{2} -\frac{k}{2}w_+ -j \right)\oao{a}{b}
 (\tau, 0 ) \ 
\bar{ch}_d \left(j,-\frac{N}{2} -\frac{k}{2} w_+ -j \right) 
\oao{a}{b} (\bar \tau, 0 ) 
\nonumber\\
\label{unextpf}
\end{eqnarray}

The extended characters of the $N=2$ algebra were introduced 
in~\cite{Eguchi:2003ik} because of their nice modular properties. 
Their decomposition in terms of bosonic coset characters 
was given in~\cite{Israel:2004xj}:\footnote{From now we assume 
that the level $k$ is integer} 
\begin{eqnarray}
\mathrm{Ch} \left( j,r \right)\oao{a}{b} & = &
\sum_{n' \in \mathbb{Z}} ch_d (j,r +kn') \oao{a}{b} (\tau,\nu ) 
\nonumber\\
&=&\sum_{n,n' \in \mathbb{Z}} 
e^{i\pi b \left(n + \frac{a}{2}\right)}
\lambda^{j}_{\frac{k}{2} Q_m - n - \frac{a}{2} -j+ kn'}  
 \  \frac{q^{\frac{k}{2(k+2)} \left( 
n + \frac{a}{2} + Q_m +2n' \right)^2}}{\eta (\tau )} 
z^{n + \frac{a}{2} + Q_m + 2n'}\nonumber\\
\end{eqnarray}
They differ from the standard $N=2$ characters, eq.~(\ref{unext2}) 
by a sum over even $N=2$ charges. 
Now, starting from eq.~(\ref{unextpf}), we define: $N=e+kp$.  
The charge $e$ is defined modulo $k$. We obtain then: 
\begin{eqnarray}
Z_{D} &=& \frac{1}{2} 
\sum_{a,b \in \zi_2}
\sum_{2j=1}^{k+1} \Upsilon (2j,1  ) 
\sum_{e \in \zi_{k}}
\sum_{p,w_+ \in \zi}  
ch_d \left(j,\frac{e}{2} +\frac{k}{2}(p-w_+)-j \right)\oao{a}{b}
 \nonumber \\ &&\qquad \qquad  \times\ \   
\bar{ch}_d \left(j,-\frac{e}{2} -\frac{k}{2}(p+w_+)-j \right) \oao{a}{b}
\nonumber \\
&=& \frac{1}{2 }
\sum_{a,b \in \zi_2}
\sum_{2j=1}^{k+1} \Upsilon (2j,1  ) 
\sum_{e \in \zi_{k}} \nonumber \\
&& \times \ \left\{ \sum_{n,\bar{n} \ \mathrm{even}} 
ch_d \left(j,\frac{e}{2} +\frac{kn}{2}-j \right)\oao{a}{b}   
\bar{ch}_d \left(j,-\frac{e}{2} -\frac{k\bar{n}}{2}-j \right) \oao{a}{b}
\right. \nonumber \\
&& \quad + \quad \left. \sum_{n,\bar{n} \ \mathrm{odd}} 
ch_d \left(j,\frac{e}{2} +\frac{kn}{2}-j \right)\oao{a}{b}   
\bar{ch}_d \left(j,-\frac{e}{2} -\frac{k\bar{n}}{2}-j \right) \oao{a}{b}
\right\}
\end{eqnarray}
So we get finally:
\begin{eqnarray}
Z_D = \frac{1}{2} 
\sum_{a,b \in \zi_2}\ 
\sum_{2j=1}^{k+1} \Upsilon (2j,1  ) \ 
\sum_{e \in \zi_{k}} \left\{ 
Ch_d \left(j,\frac{e}{2}-j \right)\oao{a}{b}
\ \bar{Ch}_d \left(j,-\frac{e}{2}-j \right) \oao{a}{b}\right. \nonumber \\
+ \left. Ch_d \left(j,\frac{e+k}{2}-j \right)\oao{a}{b}
\ \bar{Ch}_d \left(j,-\frac{e+k}{2}-j \right) \oao{a}{b}
\right\}\nonumber\\
=\frac{1}{2} \sum_{a,b \in \zi_2}\ 
\sum_{2j=1}^{k+1} \Upsilon (2j,1  ) \ 
\sum_{2g \in \zi_{2k}} 
Ch_d \left(j,g-j \right)\oao{a}{b}
\ \bar{Ch}_d \left(j,-g-j \right) \oao{a}{b},
\nonumber \\
\label{discextdec}
\end{eqnarray}
where the charge $2g$ is defined modulo $2k$. 
In short, we have proven that, for $k$ integer, the  
discrete spectrum of the supersymmetric coset $\slr / U(1)$ can 
be decomposed in terms of extended discrete characters of the 
$N=2$ algebra.


\subsection{The continuous representations}
Now we would like to identify the contributions coming from 
the continuous representations of $\slr$. These representations 
are labeled by the Casimir $c_2 = s^2 + 1/4$ with $s \in \mathbb{R}_+$, 
and their parity $2\alpha \in \zi_2$. These representations are 
neither highest nor lowest weight, and the spectrum of the 
elliptic operator is: $m=\alpha + n$, $n \in \zi$. We have 
flowed representations as well. 

We start again from eq.~(\ref{partco}). After extracting 
the poles corresponding to the discrete spectra, we are left with 
the following divergent expression: 
\begin{eqnarray}
Z_{C} = \frac{1}{\pi \eta^3 \bar{\eta}^3} 
\frac{1}{2} \sum_{a,b \in \mathbb{Z}_2} P \int_{-\infty}^{\infty} ds\ 
(q\bar{q})^{\frac{s^2}{k}}
\sum_{N,w_+ ,\tf,\tbf \in \mathbb{Z}} e^{i\pi (\tf-\tbf)b}
 \sum_{r,\bar{r} \in \mathbb{Z}} 
\delta_{r-\bar{r}+\tf-\tbf,N}\nonumber\\
\times \ \left\{ \ q^{\frac{1}{4(k+2)} \left( N-2\tf -a - kw_+ \right)^2 
+ \frac{k}{2(k+2)} \left( \tf + \frac{a}{2} + 
\frac{N-kw_+}{k} \right)^2}\right. \nonumber\\
\times \ \bar{q}^{\frac{1}{4(k+2)} \left( N+2\tbf +a + kw_+ \right)^2 
+ \frac{k}{2(k+2)} \left( \tbf + \frac{a}{2} + 
\frac{-N-kw_+}{k} \right)^2}\nonumber\\
\times \ \frac{S_r \bar{S}_{\bar{r}}}{2is+1+r+\bar{r}+ 
\tf + \tbf + a+kw_+} 
\nonumber \\
-  \ q^{\frac{1}{4(k+2)} \left( N-2(\tf +1) -a - k(w_+ + 1)\right)^2 
+ \frac{k}{2(k+2)} \left( \tf +1 + \frac{a}{2} + 
\frac{N-k(w_+ +1)}{k} \right)^2+r}\nonumber \\
\times \ \bar{q}^{\frac{1}{4(k+2)} \left( N+2(\tbf+1) +a + k(w_+ + 1)\right)^2 
+ \frac{k}{2(k+2)} \left( \tbf +1 + \frac{a}{2} + 
\frac{-N-k(w_+ +1)}{k} \right)^2+ \bar{r}}\nonumber\\
\times \ \left. \frac{S_r \bar{S}_{\bar{r}}}{2is+1+r+\bar{r}+ 
\tf + \tbf + a+k(w_+ +1)} \right\} 
\nonumber \\
\end{eqnarray}
Shifting in the second term $w_+ + 1 \to w_+$, 
$\tf + 1 \to \tf$, $\tbf + 1 \to \tbf$ (which is related
to performing spectral flow by one unit 
in both the bosonic and the fermionic sector) and using the following identity:
\begin{equation}
q^r S_r (\tau ) = S_{-r} (\tau ),
\end{equation}
we obtain finally the expression:
\begin{eqnarray}
Z_{C} &=& \frac{1}{\pi \eta^3 \bar{\eta}^3} 
\frac{1}{2} \sum_{a,b \in \mathbb{Z}_2} P \int_{-\infty}^{\infty} ds\ 
(q\bar{q})^{\frac{s^2}{k}}
\sum_{N,w_+ ,\tf,\tbf \in \mathbb{Z}} e^{i\pi (\tf-\tbf)b}\nonumber \\
&& \times \ q^{\frac{1}{4(k+2)} \left( N-2\tf -a - kw_+ \right)^2}
q^{\frac{k}{2(k+2)} \left( \tf + \frac{a}{2} + 
\frac{N-kw_+}{k} \right)^2}\nonumber \\
&& \times \ \bar{q}^{\frac{1}{4(k+2)} \left( N+2\tbf + a + kw_+ \right)^2}
\bar{q}^{\frac{k}{2(k+2)} \left( \tbf + \frac{a}{2} + 
\frac{-N-kw_+}{k} \right)^2} \sum_{r,\bar{r} \in \mathbb{Z}} 
\delta_{r-\bar{r}+\tf-\tbf,N} \left( S_r \bar{S}_{\bar{r}}
- S_{-r-1}  \bar{S}_{-\bar{r}-1} \right)
\nonumber \\
&&\times \  \frac{1}{2is+1+r+\bar{r}+ \tf + \tbf + a + kw_+}
\nonumber\\
\label{partfctcont}
\end{eqnarray}
Then, to proceed, we note that we have the following identity:
\begin{equation}
S_r + S_{-r-1} = 1,
\end{equation}
that expresses the splitting of an $s=0$ (and $\alpha=1/2$) continuous representation 
in two discrete representations. Using this expression we 
recast the last factor of eq.~(\ref{partfctcont}) in the following form:
\begin{eqnarray}
\sum_{r\in \zi} (S_r -1/2) 
\frac{1}{2is +  1 + 2r + 2\tf + a - 
N + k w_+}\nonumber\\ 
+ \sum_{\bar{r}\in \zi} (\bar{S}_{\bar{r}} -1/2) 
\frac{1}{2is +  1 + 2\bar{r} + 2\tbf + a
+N + k w_+} 
\nonumber \\
\label{firstsplit}
\end{eqnarray}
First we would like to isolate the divergent term from this expression. 
By expanding $\vartheta \oao{a}{b} (\tau,\nu )/\vartheta_1 (\tau ,\nu)$
in powers of $e^{2i\pi\nu}$, 
we can also prove the following generalized identity:
\begin{equation}
\sum_{n} (-)^{b(n+a/2)} q^{\frac{1}{2} (n+a/2)^2}S_{N-n} 
+ (-)^{ab} \sum_{n} (-)^{b(n+a/2)} q^{\frac{1}{2} (n+a/2)^2}S_{-N-1-a-n}
= \vartheta \oao{a}{b} ( \tau , 0 )
\end{equation}
This allows to split naturally the left-moving and right-moving 
contributions of eq.~(\ref{firstsplit}) in sums over positive integers 
as follows:
\begin{eqnarray}
\sum_{r=0}^{\infty} \left[
\frac{\left(S_{r-\tf}-1 \right)+1/2}{1+2r+2is+a-N+kw_+} 
+ \frac{(-)^{ab} \left( S_{r-\tf-a}-1 \right)+1/2}{1+2r-2is-a+N-kw_+}
\right]\nonumber \\
+ \sum_{\bar{r}=0}^{\infty} \left[
\frac{\left(\bar{S}_{\bar{r}-\tbf}-1\right)+1/2}{1+2\bar{r}+2is+a+N+kw_+} 
+ \frac{(-)^{ab} \left( \bar{S}_{\bar{r}-\tbf-a}-1 \right)+1/2}
{1+2\bar{r}-2is-a-N-kw_+}
\right]\nonumber \\
\label{lastsplit}
\end{eqnarray}
One could have obtained naturally this decomposition by expanding directly 
the supersymmetric determinant $|\vartheta \oao{a}{b} / \vartheta_1 |^2$ 
in powers of the holonomies from the beginning of the computation.
The first terms of the expansion of the function $S_r$, eq.~(\ref{srfct}) 
are:
$$S_r = 1 - e^{-2\pi \tau_2 (r+1)}e^{2i\pi \tau_1 (r+1)} 
+ \mathcal{O} \left( e^{-2\pi \tau_2 (2r+3 )}\right) $$
Therefore, as $\tau_2 >0$ in the fundamental domain,  
all the terms in the expansion of $S_{r - \tf} -1$ 
are sub-leading with respect to the constant $1/2$ term in the sum 
over $r \in \mathbb{N}$. 
This leading constant term gives a 
logarithmic divergence that needs to be correctly regularized. 
Since this term does not depend on the modular parameter $\tau$ it 
can be interpreted as a density of states. 

\subsubsection*{The density of long string states}
The procedure to regularize this divergent factor was already 
discussed in~\cite{Maldacena:2000kv} for thermal $AdS_3$ and 
in a closer context in~\cite{Hanany:2002ev} for the bosonic coset.
We adopt the same ``Liouville wall'' regularization as in the aforementioned 
papers and find contributions like: 
\begin{equation}
\sum_{n=0}^{\infty} \frac{e^{-n\epsilon}}{A+n} 
= \log \epsilon + \frac{d}{dA} \log \Gamma (A) 
+ \mathcal{O} (\epsilon )
\end{equation}
By using this regularization procedure for the four divergent pieces 
of~(\ref{lastsplit}) and adding the contributions 
of $s>0$ and $s<0$ from the integral, we obtain:
\begin{eqnarray}
\rho (s;N,w_+;a) &=&
\frac{1}{\pi} \log \epsilon + \nonumber\\
&&\frac{1}{8i\pi} \frac{d}{ds} \log \left\{ 
\frac{\Gamma \left( \frac{1}{2}-is+\frac{a}{2}-\frac{N}{2}
+\frac{k}{2}w_+ \right)}{\Gamma \left( \frac{1}{2}+is+\frac{a}{2}-\frac{N}{2}
+\frac{k}{2}w_+ \right) }
\frac{\Gamma \left( \frac{1}{2}-is-\frac{a}{2}+\frac{N}{2}
-\frac{k}{2}w_+ \right)}{\Gamma \left( \frac{1}{2}+is-\frac{a}{2}+\frac{N}{2}
-\frac{k}{2}w_+ \right) }\right.\nonumber \\ 
&&\qquad \quad \times \qquad 
\frac{\Gamma \left( \frac{1}{2}-is+\frac{a}{2}+\frac{N}{2}
+\frac{k}{2}w_+ \right)}{\Gamma \left( \frac{1}{2}+is+\frac{a}{2}+\frac{N}{2}
+\frac{k}{2}w_+ \right) }
\left. \frac{\Gamma \left( \frac{1}{2}-is-\frac{a}{2}-\frac{N}{2}
-\frac{k}{2}w_+ \right)}{\Gamma \left( \frac{1}{2}+is-\frac{a}{2}-\frac{N}{2}
-\frac{k}{2}w_+ \right) }  \right\}
\nonumber \\
\end{eqnarray}
The divergent factor in the first line has been argued 
in~\cite{Maldacena:2000kv} to correspond to the divergence due 
to the infinite volume available for the long strings. The second one gives 
the regularized density of continuous representations. 
Thinking of the duality of the supersymmetric $\slr / U(1)$
coset with $N=2$ Liouville, we can view this result in terms of the ratio of 
the reflection amplitudes in two $N=2$ Liouville theories differing 
by a flipping of sign in the exponential potential. In addition 
to the usual wall of N=2 Liouville, we add another opposite one to prevent the 
long strings from having an infinite volume at their disposal.  
In fact, our result matches the computations of two-point functions 
in N=2 Super-Liouville\footnote{These computations correspond 
to the sector $\tmt=\tbmt$.}, both in the NS and R sectors~\cite{Ahn:2003tt}.
This is a method preserving $N=2$ superconformal symmetry 
to confine the strings to a finite volume and obtain a regularized 
density of states.\footnote{Note that the potential used to regularize the volume
divergence is largely arbitrary and each choice of regularization would lead
to a different regularized density of states. In our case, preservation of the
extended $N=2$ superconformal symmetry severely limits the choice of
regularization.} If we use the parameterization~(\ref{paramq}) for the eigenvalues
of the total currents $\mj^3$ and $\bar{\mj}^3$, this expression is just:
\begin{eqnarray}
\rho (s;\tmt,\tbmt;a) &=& \frac{1}{\pi} \log \epsilon 
+\frac{1}{8i\pi} \frac{d}{ds} \log \left\{ 
\frac{\Gamma \left( \frac{1}{2}-is+\frac{a}{2}-\tmt \right)}
{\Gamma \left( \frac{1}{2}+is+\frac{a}{2}-\tmt \right) }
\frac{\Gamma \left( \frac{1}{2}-is-\frac{a}{2}+\tmt \right)}
{\Gamma \left( \frac{1}{2}+is-\frac{a}{2}+\tmt \right) }\right.\nonumber \\ 
&&\qquad \qquad \qquad \times \qquad \quad 
\frac{\Gamma \left( \frac{1}{2}-is+\frac{a}{2} -\tbmt \right)}
{\Gamma \left( \frac{1}{2}+is+\frac{a}{2}-\tbmt \right) }
\left. \frac{\Gamma \left( \frac{1}{2}-is-\frac{a}{2}+\tbmt \right)}
{\Gamma \left( \frac{1}{2}+is-\frac{a}{2}+\tbmt \right) }  \right\}
\nonumber \\
&=&  \frac{1}{\pi} \log \epsilon 
+\frac{1}{4i\pi} \frac{d}{ds} \log \left\{ 
\frac{\Gamma \left( \frac{1}{2}-is+\frac{a}{2}-\tmt \right)}
{\Gamma \left( \frac{1}{2}+is+\frac{a}{2}-\tmt \right) }
\frac{\Gamma \left( \frac{1}{2}-is-\frac{a}{2}+\tbmt \right)}
{\Gamma \left( \frac{1}{2}+is-\frac{a}{2}+\tbmt \right) }  \right\}
\nonumber \\
\label{density}
\end{eqnarray}
where we have used in the last line the fact that 
$\tmt - \tbmt \in \zi$.
To summarize, we have obtained the continuous spectrum 
in the following form: 
\begin{eqnarray}
Z_C &=&  
\frac{1}{2}\sum_{a,b \in \zi_2} \sum_{N,w_+ \in \zi}
\int_{0}^{\infty} ds \ \rho (s;N,w_+;a)  
ch_c \left(s,\frac{N}{2}-\frac{k}{2}w_+\right) \oao{a}{b} 
\bar{ch}_c \left( s,-\frac{N}{2}-\frac{k}{2}w_+ \right) \oao{a}{b} 
\nonumber \\
\end{eqnarray} 
We have introduced the following continuous characters of the $N=2$ 
algebra (with $2m \in \zi$):
\begin{eqnarray}
ch_c (s,m) \oao{a}{b} (\tau, \nu ) &=& \frac{1}{\eta^3}
\ q^{\frac{s^2 + m^2}{k}} z^{\frac{2m}{k}}
\vartheta \oao{a}{b} (\tau , \nu ) \nonumber \\
&=& \frac{1}{\eta} \sum_n e^{i\pi b \left(n+ \frac{a}{2} \right)}\ 
z^{\frac{2m}{k}+n+\frac{a}{2}} \
q^{\frac{k}{2(k+2)} \left(n+\frac{a}{2} + \frac{2m}{k}\right)^2}
\hat{\lambda}_{1/2+is,n+\frac{a}{2}-m} (\tau )\nonumber \\
\end{eqnarray}
In the last line, we have decomposed the continuous $N=2$ characters in terms 
of the coset characters for the continuous representations:
\begin{equation}
\hat{\lambda}_{1/2+is,m}(\tau ) = \frac{1}{\eta^2} \ q^{\frac{s^2}{k} +
\frac{m^2}{k+2}}
\end{equation}
\subsubsection*{Decomposition in extended continuous N=2 characters}
We have already shown that the discrete characters of the 
supersymmetric coset can be decomposed into extended discrete 
characters of $N=2$. Now we would like to introduce the 
{\it extended continuous} characters of $N=2$~\cite{Eguchi:2003ik}. 
They are also given by (see~\cite{Israel:2004xj} for details):
\begin{eqnarray}
 Ch_c (s,m) \oao{a}{b} (\tau , \nu ) 
&=& \sum_{n \in \zi} 
ch_c (s,m+kn) \oao{a}{b} (\tau , \nu ) \nonumber \\
&=& q^{\frac{s^2}{k}} 
\frac{\vartheta \oao{a}{b} (\tau , \nu )}{\eta (\tau )}
\Theta_{2m,k} \left(\tau, \frac{2\nu}{k}\right)
\label{extmassiv}
\end{eqnarray}
where we have introduced the classical theta function at level $k$: 
\begin{equation}
\Theta_{m,k} (\tau , \nu ) = 
\sum_{n \in \zi} q^{k \left( n+ \frac{m}{2k} \right)^2}
z^{k \left( n+ \frac{m}{2k} \right)}.
\end{equation}
To decompose the partition function in terms of extended characters we 
need another regulator compatible with them. In fact, the density of 
continuous representations we obtained before, eq.~(\ref{density}), 
is not invariant under spectral flow shifts by $k$ units. To obtain a invariant 
regulator, we could  have started our computation with the expansion of the
full determinant 
$|\vartheta \oao{a}{b} / \vartheta_1 |^2 \ \Theta_{m,k} \bar{\Theta}_{-m,k}$
in powers of the holonomies. Then the sum over $r$ can be splitted using the 
the following identity:
\begin{eqnarray}
\sum_{n,n'} (-)^{b(n+a/2)} q^{\frac{1}{2} (n+a/2)^2 + 
k\left( n' + \frac{m}{2k} \right)^2} S_{N-n-kn'}
\nonumber \\ 
+ (-)^{ab} \sum_{n,n'} (-)^{b(n+a/2)} 
q^{\frac{1}{2} (n+a/2)^2 + k\left( n' - \frac{m}{2k} \right)^2
}S_{-N-1-a-m-n-kn'}
\nonumber \\
= \vartheta \oao{a}{b} ( \tau , 0 ) \Theta_{m,k} (\tau ,0). 
\end{eqnarray}
Now we can split the variable $N$ as for the discrete representations: 
$N = kp +e$, with $e \in \zi_{k}$, and then we define the mod $2k$ 
charge $2g=e + kt$, $ t= 0,1$.  We obtain the following 
anti-symmetric combination of extended continuous $N=2$ characters:
\begin{eqnarray}
Z_C = \frac{1}{2} \sum_{a,b \in \zi_2}
 \int_{0}^{\infty} ds \   
\sum_{2g \in \zi_{2k}} \ \rho (s,g,-g;a)\ 
Ch_c \left(s, g \right) \oao{a}{b} \ 
\times \ 
\bar{Ch}_c \left( s,-g \right) \oao{a}{b} 
\nonumber \\
\end{eqnarray} 
with the following density of states:
\begin{eqnarray}
\rho (s,g,\bar{g};a) = \frac{1}{\pi} \log \epsilon 
+\frac{1}{4i\pi} \frac{d}{ds} \log \left\{ 
\frac{\Gamma \left( \frac{1}{2}-is+\frac{a}{2}-g \right)}
{\Gamma \left( \frac{1}{2}+is+\frac{a}{2}-g \right) }
\frac{\Gamma \left( \frac{1}{2}-is-\frac{a}{2} +\bar{g} \right)}
{\Gamma \left( \frac{1}{2}+is-\frac{a}{2}+\bar{g} \right) }
\right\}
\nonumber \\
\label{densityext}
\end{eqnarray}
Note that the leading divergent term is modular invariant 
by itself, as we now from~\cite{Eguchi:2003ik} 
and~\cite{Israel:2004xj}. We have shown here that 
different regularization procedures give 
different finite parts for the density of states.
Finally we would like to stress that,
due to the expression of the continuous characters, the leading
contribution to the partition function is given by a free field
computation. 


\subsection{Puzzling sectors of states}
In the last subsections we have succeeded in decomposing the partition 
function in both discrete and continuous representations of the 
supersymmetric coset theory $\slr / U(1)$. The contribution 
of the continuous representations came with an infrared divergence, 
that we regularized.
The regularized expression  gives a non-trivial density of states.
However, the partition function contains also a piece 
corresponding to the the finite part of the
partition function after the extraction of the discrete representations.
These new sectors are explicitely given by:
\begin{eqnarray}
\left[ \sum_{r \in \zi} 
\frac{S_r}{\frac{1}{2}+is+\frac{a}{2}-\tmt+r} 
- \sum_{r \in \mathbb{N}} \frac{1}{\frac{1}{2}+is+\frac{a}{2}-\tmt+r}
\right] \frac{q^{\frac{s^2+\tmt^2}{k}}}{\eta^3} 
\vartheta \oao{a}{b},
\end{eqnarray}
and we have a similar term with non-trivial contribution from the 
right-moving sector. These terms cannot be interpreted as 
coming from usual highest weight representations of 
$\slr$. Our regularization procedure is not obviously modular 
invariant and/or compatible with the $\slr$ affine 
symmetry of the parent theory. 
A modular invariant regularization is provided by using for the $H_3$ determinant 
the function: $1/|\vartheta_1 (\nu | \tau )|^{2(1-\epsilon)}$, with $\epsilon >0$. 
However the explicit integration over the holonomies becomes 
challenging.


\subsection{Comments on the Witten index}
The twisted RR sector of the partition function:
Tr$_{RR} \ (-)^F q^{L_0} (-)^{\bar{F}} q^{\bar{L}_0}$ is a constant 
-- the Witten index~\cite{Witten:df} -- 
and we observe that it can be computed straightforwardly, using the 
Lagrangian form of the un-decomposed partition function, eq.~(\ref{lagr}): 
\begin{eqnarray}
I_W &=& k 
\int_{0}^{1} ds_1 \ ds_2 \
\frac{1}{\left|
\vartheta \oao{1+2s_1}{1+2s_2} \right|^2} 
\sum_{M,w_{+} \in \mathbb{Z}}
e^{-\frac{k\pi}{\tau_2} \left|(w_+ + s_1 )\tau 
-(M+s_2)\right|^2} \left| \vartheta \oao{1+2s_1}{1+2s_2} \right|^2 
\nonumber \\
&=&   k 
\int_{\mathbb{R}^2} du_1 \ du_2 \ e^{-\frac{k\pi}{\tau_2}
  \left|u_1 \tau - u_2\right|^2} = 1
\label{lagwitt}
\end{eqnarray}
Even if the partition function for the super-coset, eq.~(\ref{partco}) 
is divergent, as the bosonic coset partition 
function~\cite{Hanany:2002ev}, the Witten index is finite.   
However the computation is very sensitive to the normalization 
of the path integral that lead to the expression~(\ref{partsl2}). 
We will explain now how the present computation is consistent with the decomposition 
of the supersymmetric coset partition function into $N=2$ characters.
\paragraph{Contribution of the extended discrete characters:}
we recall that the extended discrete characters of $N=2$ are given 
by:
\begin{equation}
Ch_d (j,r) \oao{a}{b} (\tau,\nu ) =
\frac{1}{\eta^3 (\tau )} 
\sum_{n \in \zi}
q^{ \frac{-(j-1/2)^2+(j+r+a/2 + kn)^2}{k}} z^{\frac{2j+2r+a}{k}}
\frac{\vartheta \oao{a}{b} (\tau, \nu )}{1+(-)^b z\ q^{(1+a)/2+r+kn}}
\end{equation}
Therefore in the twisted Ramond sector we have: 
\begin{equation}
Ch_d (j,r) \oao{1}{1} (\tau,0 ) = 
\sum_{n \in \zi}
q^{ \frac{-(j-1/2)^2+(j+r+1/2 + kn)^2}{k}} 
\delta_{r+kn+1,0} = \delta_{r+1,0 \ \mathrm{mod} \ k}
\end{equation}
So the full contribution of the discrete part of the partition function  
to the index is:
\begin{eqnarray} 
I_D &=&   
\sum_{2j=1}^{k+1} \ \Upsilon (2j,1  ) \  
\sum_{m \in \zi_{2k}} \ 
Ch_d \left(j,\frac{m}{2}-j \right)\oao{1}{1}
\ \bar{Ch}_d \left(j,-\frac{m}{2}-j \right) \oao{1}{1} \nonumber \\
&=&  \sum_{2j=1}^{k+1} \ \Upsilon (2j,1  ) \  
\sum_{m \in \zi_{2k}} \
\delta_{m-2j,0 \ \mathrm{mod} \ 2} \ \times \ 
\delta_{m-2j+2,0  \ \mathrm{mod} \ 2k} \ \times \ 
\delta_{m+2j-2,0  \ \mathrm{mod} \ 2k}  \nonumber\\
&=& 1
\end{eqnarray}
Therefore only one primary state from the discrete representations: 
$j=1$, $m=0$ (left and right) contribute to the Witten index in the axial coset. 

The continuous representations -- and the puzzling 
sectors as well -- do not contribute to the Witten 
index. So the result is in perfect agreement with the direct computation of 
the Witten index using the Lagrangian partition function.

\section{The double scaled little string theory}
\label{sectlst}
The little string theory~\cite{Seiberg:1997zk}~\cite{Aharony:1998ub}
(LST) is the decoupling limit of the theory living on the NS five-branes, 
obtained by sending the string coupling constant to zero. This 
theory is non-gravitational but shares many properties with conventional 
string theories. A holographic description is provided by the 
near-horizon geometry of the five-branes background, which 
is an exact superconformal field 
theory~\cite{Kounnas:1990ud}\cite{Callan:dj}\cite{Antoniadis:1994sr}
based on (for $k$ five-branes) 
$SU(2)_{k} \times \mathbb{R}_{Q^2=2/k}$. However, because of the 
linear dilaton this string background is strongly coupled down the throat.
 The authors 
of~\cite{Giveon:1999px} show that this strong coupling problem can 
be avoided by going to the Higgs phase of the little string theory, 
and taking a double scaling limit. Using the duality 
between $N=2$ super-Liouville and the supersymmetric coset 
$\slr / U(1)$, they argue that this background 
(the DSLST) is given by the following orbifold:
$$\frac{\slr_{k} / U(1) \times SU(2)_{k} / U(1)}{\zi_{k}}$$ 
This worldsheet CFT also describes superstrings on a 
singular non-compact Calabi-Yau space, in this case on an 
$A_{k-1}$ ALE manifold~\cite{Ooguri:1995wj}.
Another regularization of the 
strong coupling problem is given by a null deformation of 
$AdS_3$, corresponding to adding a stack of fundamental strings 
to the five-branes setup~\cite{Israel:2003ry}.
In the following we will compute the partition function for 
the superconformal field theory of the double scaled little 
string theory, and show explicitely that 
it gives a spacetime supersymmetric string background (related 
issues were considered in~\cite{Yamaguchi:2000dz}).

\subsection{N=2 minimal models: a reminder}
Here we would like to recall the partition function of the
supersymmetric $SU(2)_k /U(1)$ coset. The building blocks are the 
N=2 minimal models characters which are given by 
(see~\cite{Kiritsis:1986rv}\cite{Dobrev:1986hq}\cite{Matsuo:1986cj}):
\begin{equation}
\chi^{j\ (s)}_{m} (\tau,\nu )= \sum_{n \in \mathbb{Z}_{k-2}} 
c^{j}_{m+4n-s} (\tau )
\Theta_{2m+k(4n-s),2k(k-2)} \left(\tau, \frac{\nu}{k} \right)
\label{MMchar}
\end{equation}
where the $c^{j}_m$ are the $SU(2)_{k-2}$ string functions. 
The charge s is defined modulo 4: $s=0,2$ are the NS sectors with 
$(-)^F = +1,-1$ and  $s=1,3$ are the R sectors with 
$(-)^F = +1,-1$. Note that in the NS sector, we have 
$2j+m$ even, while in the R sector, $2j+m$ is odd. 
We have the following symmetry:
\begin{equation}
\chi^{(k-2)/2-j\ (s+2)}_{m+k} (\tau,\nu )
= \chi^{j\ (s)}_{m} (\tau,\nu ),
\label{conjmin}
\end{equation}
and the modular transformation: 
\begin{equation}
\chi^{j\ (s)}_{m} (-1/\tau,0) = \frac{1}{k} 
\sum_{j',m',s'} 
\sin \left( \pi \frac{(2j+1)(2j'+1)}{k} \right)
e^{i\pi \left(\frac{mm'}{k} - \frac{ss'}{2} \right)}
\chi^{j'\ (s')}_{m'} (\tau, 0).
\end{equation} 
Grouping the NS and R states we obtain the RNS characters:
\begin{equation}
\mc^{j}_{m} \oao{a}{b} = 
\left( \chi^{j\ (a)}_{m} + (-)^{b} \chi^{j\
(a+2)}_{m} \right) e^{i\pi \frac{ab}{2}}
\end{equation}
with the symmetry:
\begin{equation}
\mc^{(k-2)/2-j}_{m+k} \oao{a}{b} = 
(-)^ b \mc^{j}_{m} \oao{a}{b}. 
\label{conjmin2}
\end{equation}
Then a type 0B modular invariant is given by:
\begin{equation}
Z_{MM} = \frac{1}{2} \sum_{a,b \in \mathbb{Z}_2}
\sum_{2j,2j'=0}^{k-2} \sum_{n,n' \in \mathbb{Z}_{2k}} 
N_{j j'} M^{nn'} \mc^{j}_{n} \oao{a}{b}
 \bar{\mc}^{j'}_{n'} \oao{a}{b}
\end{equation}
where $N_{j j'}$ is any modular invariant of $SU(2)$, 
for example the A-invariant:  $N_{j j'}=\delta_{j,j'}$.

For the modular invariant of the level $k$ theta functions 
$M^{mm'}$ we can choose the diagonal one $\delta_{m,m'}$. Of great 
importance for the following will be the $\zi_k$ 
orbifold modular invariant:
\begin{eqnarray*}
Z &=& \sum_{\substack{n-n' = 0 \ \mathrm{mod} \ 2k \\
n+n'= 0 \ \mathrm{mod}\  2}}
\Theta_{n,k} \bar{\Theta}_{n',k} 
= \frac{1}{k} \sum_{\gamma',\delta' \in \mathbb{Z}_k} 
\sum_{n \in \mathbb{Z}_{2k}} 
e^{2i\pi (n-\gamma')\frac{\delta'}{k}}
\Theta_{n,k} \Theta_{-n+2\gamma',k}\\
\end{eqnarray*}
Then we have the following orbifold sectors:
\begin{equation}
Z_{MM} \oao{a;a}{b;b} \oao{\gamma'}{\delta'} = 
\sum_{2j=0}^{k-2} \sum_{m \in \mathbb{Z}_{2k}} 
\mc^{j}_{m} \oao{a}{b}
 \bar{\mc}^{j'}_{-m+2\gamma'} \oao{a}{b} 
e^{2i\pi (m-\gamma' )\frac{\delta'}{k}}
\label{orbminmod}
\end{equation}


\subsection{Combining the two cosets}
From now on we consider the two coset theories together, 
$\slr / U(1)$ and $SU(2) / U(1)$ with the same supersymmetric 
level $k$. The central charges add up to $c=6$:
\begin{equation}
c = c_{\slr /U(1),\ susy}+ c_{SU(2) / U(1),\ susy} 
= \frac{3(k+2)}{k} + \frac{3(k-2)}{k} = 6.
\end{equation} 
In order to obtain eventually a spacetime supersymmetric background, 
we have to consider a $\zi_k$ orbifold of this superconformal 
field theory. 
\subsubsection*{N=2 charges}
Let us begin with the $\slr / U(1)$ part. 
The N=2 charges of the twisted 2D black hole are, from
eq.~(\ref{partcotw}):
\begin{equation}
\mq_L = \tf+\frac{a}{2}+\frac{N-\gamma-kw_+}{k} , \ \mq_{R} =
\tbf +\frac{a}{2}-\frac{N+\gamma+kw_+}{k},
\end{equation}
combined with a projector: $e^{2i\pi N \delta /k}$.\\
Now let us consider the $SU(2)/ U(1)$ factor. From eq.~(\ref{orbminmod}), 
The N=2 charges of the twisted $N=2$ minimal model are:
\begin{equation}
\mq' _L = -s/2+\frac{m}{k} \mod 2, \ \mq' _{R} = 
-s/2 +\frac{-m+2\gamma'}{k}\ \mod 2 
\end{equation}
combined with a projector: $e^{2i\pi (m - \gamma' )\delta' /k}$.\\
\boldmath
\subsubsection*{The $\zi_k$ orbifold}
\unboldmath
If we choose to identify the two projections (i.e. we mod out by 
a diagonal $\zi_k$), namely
$\gamma' = \gamma$, $\delta' = \delta$, we find the
following set of total N=2 charges:
\begin{eqnarray}
\mq^{tot}_L &=& \tf +\frac{a}{2}- \frac{s}{2}+\frac{m+N-\gamma}{k} 
-w_+ \mod 2 \nonumber \\ 
\mq^{tot}_R &=& \tbf +\frac{a}{2} - \frac{s}{2} 
-\frac{N+m-\gamma}{k} -w_+ \mod 2
\end{eqnarray}
with the projector: $$e^{2i\pi \left(m-\gamma+N \right) \frac{\delta}{k}}.$$
Summing over $\delta$ gives  the constraint 
\begin{equation} m-\gamma+N = 0 \ \mathrm{mod}\ k
:=  kp,
\end{equation}
such that the left and right $N=2$ charges are both integer.
So the type 0B partition function for this theory is: 
\begin{equation}
Z_{\slr / U(1) \times SU(2) / U(1)}^{(0B)} 
= \frac{1}{2} \sum_{a,b \in \zi_2} 
\frac{1}{2k} \sum_{\gamma,\delta \in \mathbb{Z}_{k}} Z_{BH} \oao{a;a}{b;b} 
\oao{\gamma}{\delta} Z_{MM} \oao{a;a}{b;b} \oao{\gamma}{\delta}
\label{partdlst0}
\end{equation}
Then we conclude that the $\zi_k$ orbifold of the 
superconformal theory $\slr_k / U(1) \times SU(2)_k / U(1)$ 
has an integer spectrum of $N=2$ R-charges. It will then be 
possible to construct a spacetime supersymmetric vacua 
using the spectral flow of the $N=2$ algebra~\cite{Gepner:1989gr}.
Now we consider the complete superstring background for 
the doubly scaled little string theory, i.e. 
the orbifold of  $\slr_k / U(1) \times SU(2)_k / U(1)$  
times six dimensional Minkowski space. In the light-cone gauge 
we obtain the following partition function for type $0B$ 
DSLST:
\begin{equation}
Z_{DSLST}^{(0B)} 
= \frac{1}{2} \sum_{a,b \in \zi_2} 
\frac{1}{2k} \sum_{\gamma,\delta \in \mathbb{Z}_{k}} Z_{BH} \oao{a;a}{b;b} 
\oao{\gamma}{\delta} Z_{MM} \oao{a;a}{b;b} \oao{\gamma}{\delta}
\frac{1}{(8\pi^2 \tau_2 )^2 \eta^4 \bar{\eta}^4}
\frac{\vartheta^2 \oao{a}{b}}{\eta^2} \frac{\bar \vartheta^2 \oao{a}{b}}{
\bar \eta^2}
\label{partdlstc0}
\end{equation}

\subsection{N=4 structure of the double scaled little string theory}
In this section we would like to study the underlying $N=4$ structure 
of the theory, and in particular relate the characters we obtained 
to the $N=4$ characters of the near-horizon geometry of coinciding NS5-branes 
-- the $SU(2)_k \times \mathbb{R}_Q$ SCFT.

\subsubsection*{The N=4 algebra}
The superconformal symmetry of the theory 
$\slr / U(1) \times SU(2) / U(1)$ is enlarged to a small 
N=4 superconformal algebra, due to the $\zi_k$ orbifold.
Explicitely, we have the following supercurrents and 
$SU(2)_1$ currents~\cite{Kounnas:1993ix}:
\begin{eqnarray}
G^{\pm}  & = & - \sqrt{\frac{2}{k}}
\left[ \sqrt{k+2} \, \pi_{1}^{-} 
e^{-i\sqrt{\frac{k-2}{2(k+2)}} H^-} 
+  \sqrt{k-2} \, \psi_{1}^{-} e^{i\sqrt{\frac{k+2}{2(k-2)}} H^-} 
\right] e^{\pm \frac{i}{\sqrt{2}} H^+} \nonumber \\
\tilde{G}^{\pm} & = &  \hphantom{-} \sqrt{\frac{2}{k}}
\left[ \sqrt{k+2}\,  
\pi_{1}^{+} e^{i\sqrt{\frac{k-2}{2(k+2)}} H^-} 
-  \, \sqrt{k-2} \psi_{1}^{+} e^{-i\sqrt{\frac{k+2}{2(k-2)}} H^-} 
\right] 
 e^{\pm \frac{i}{\sqrt{2}} H^+} \nonumber \\
(S^3, S^{\pm})  & = & \hphantom{-} \left(\frac{i}{\sqrt{2}} \partial H^+ , 
e^{\pm i\sqrt{2} H^+}\right) 
\label{DLSTalg}
\end{eqnarray}
Here $\psi_{1}^{\pm}$ are the first parafermionic currents of $SU(2)
/U(1)$, $\pi_{1}^{\pm}$ are those of $\slr / U(1)$ and 
$H^{\pm}$ are free bosons. The boson $H^+$ is compactified at the self-dual 
radius $R = \sqrt{2}$ -- in order to have an $SU(2)_1$ R-symmetry -- 
and the boson $H^-$ is compactified at the radius: 
$$R_{H_-} = \sqrt{\frac{2(k+2)}{k-2}}.$$ 
To exhibit the N=4 structure in our partition function,
eq.~(\ref{partdlst0}), we use the identity:
\begin{equation}
\frac{k}{2(k+2)}\mq^2 + \frac{k}{2(k-2)} \mq'^2 = \frac{1}{4}
\left(\mq + \mq' \right)^2 + \frac{k-2}{4(k+2)} 
\left( \mq - \frac{k+2}{k-2} \mq' \right)^2
\end{equation}
After the $\zi_k$ orbifold, the charges of the first factor are  
half-integer as they should be for an $SU(2)_1$ lattice.
\subsubsection*{On the equivalence between DSLST and LST superconformal 
algebras}
We rewrite the generators of the $N=4$ algebra for 
$\slr / U(1) \times SU(2) / U(1)$, eq.~(\ref{DLSTalg}), 
with a free field representation for the $\slr / U(1)$ 
piece~\cite{Griffin:1990fg}:
\begin{eqnarray}
G^{\pm} = -\sqrt{\frac{2}{k}}\left[ 
\left( \sqrt{\frac{k+2}{2}} \partial X - i \sqrt{\frac{k}{2}} \partial \rho \right) 
e^{- i \sqrt{\frac{2}{k+2}}X  -i\sqrt{\frac{k-2}{2(k+2)}} H^-}
+  \sqrt{k-2} \, \psi^{-} e^{i\sqrt{\frac{k+2}{2(k-2)}} H^-} 
\right]  \nonumber \\ \times \quad e^{\pm \frac{i}{\sqrt{2}} H^+},
\nonumber\\
\end{eqnarray}
and we have a similar expression for the other $SU(2)$ doublet. The fields 
$X(z)$ and $\rho(z)$ are canonically normalized chiral bosons, with a 
background charge on $\rho$: $Q_{\rho} = \sqrt{2/k}$.
Now we perform the following $O(2)$ rotation between the free fields 
$X$ and $H^-$:
\begin{eqnarray}
H^- &=& \cos \theta \ \hat{H}^{-} - \sin \theta \ \hat{X}  \nonumber \\
X\hphantom{^-}&=& \cos \theta \ \hat{X} + \sin \theta \ \
\hat{H}^- \, , \qquad
\mathrm{with} \ \cos^2 \theta = \frac{k-2}{k+2}.
\end{eqnarray}
And one obtains the following supercurrents:
\begin{eqnarray}
\hat{G}^{\pm} = -\sqrt{\frac{2}{k}}\left[ 
\left( \sqrt{\frac{k-2}{2}} \partial \hat{X} + \sqrt{2} \partial \hat{H}^- 
- i \sqrt{\frac{k}{2}} \partial \rho \right) 
e^{-\frac{i}{\sqrt{2}} \hat{H}^-}
+  \sqrt{k-2} \, \psi^{-} e^{-i \sqrt{\frac{2}{k-2}} \hat{X} 
+ \frac{i}{\sqrt{2}} \hat{H}^-} 
\right] \nonumber \\ \times \quad e^{\pm \frac{i}{\sqrt{2}} H^+},
\nonumber\\
\end{eqnarray}
Now, if we introduce the purely bosonic current 
of the $SU(2)_{k-2}$ Cartan sub-algebra: 
\begin{equation}
I^3 (z) = i\sqrt{\frac{k-2}{2}} \partial \hat{X},
\end{equation}
One obtains exactly the N=4 algebra for 
$SU(2)_k \times \mathbb{R}_{Q^2=2/k}$~\cite{Kounnas:1993ix}, i.e. 
the near-horizon limit of coincident five-branes background. 
The theory differs only by the $N=2$ super-Liouville potential, 
that breaks the $SO(4)$ R-symmetry 
of the LST (the $SU(2)_{k-1} \times SU(2)_{k-1}$ isospin of the 
N=4 algebra) to the $U(1) \times \zi_k$ symmetry of the DSLST. 
\subsubsection*{Matching the continuous spectra}
Implementing this isomorphism, we expect that the continuous 
spectrum, that correspond to asymptotic states, will be the same 
in $\slr / U(1) \times SU(2) / U(1)$ -- the doubly scaled LST -- and in
$SU(2) \times \mathbb{R}_Q$ -- the LST. Explicitely we have 
for the $\slr / U(1) \times SU(2) / U(1)$
(still in type 0B):
\begin{eqnarray}
Z_C = \frac{1}{4k} \sum_{a,b \in \zi_2}
  \   \sum_{\gamma, \delta \in \zi_k}
\int_{0}^{\infty} ds \ 
\sum_{2g \in \zi_{2k}} e^{4i\pi g \frac{\delta}{k}}
\ \rho (s,g;\gamma;a)\ 
Ch_c \left(s, g-\gamma/2 \right) \oao{a}{b} \ \nonumber \\
\times \  \bar{Ch}_c \left( s,-g- \gamma/2 \right) \oao{a}{b} 
\ \sum_{2j=0}^{k-2} \sum_{m \in \mathbb{Z}_{2k}} 
\mc^{j}_{m} \oao{a}{b}
 \bar{\mc}^{j}_{-m+2\gamma} \oao{a}{b} 
e^{2i\pi (m-\gamma)\frac{\delta}{k}}
\end{eqnarray}  
With the density of states:
\begin{eqnarray}
\rho (s,g;\gamma;a) = \frac{1}{\pi} \log \epsilon 
+\frac{1}{4i\pi} \frac{d}{ds} \log \left\{ 
\frac{\Gamma \left( \frac{1}{2}-is+\frac{a}{2}-g+\frac{\gamma}{2} \right)}
{\Gamma \left( \frac{1}{2}+is+\frac{a}{2}-g +\frac{\gamma}{2} \right) }
\frac{\Gamma \left( \frac{1}{2}-is-\frac{a}{2} -g - \frac{\gamma}{2} \right)}
{\Gamma \left( \frac{1}{2}+is-\frac{a}{2}+g - \frac{\gamma}{2} \right) }
\right\}
\nonumber \\
\label{densityexttw}
\end{eqnarray}
We should only compare the leading divergent part of the 
density of states to match the spectra of the two theories, because 
the regularized density depends on the phase shift of the $N=2$
super-Liouville potential
which we used to regularize the partition function.
For the divergent part, we obtain then, after solving for the constraints:
\begin{eqnarray}
\frac{\log \epsilon}{4k\pi} \sum_{a,b \in \zi_2}
  \   \sum_{\gamma, \delta \in \zi_k} \sum_{2j=0}^{k-2}
\int_{0}^{\infty} ds \ 
\sum_{m \in \zi_{2k}} \nonumber \\
\times \ \left\{ \left( Ch_c \left(s, -\frac{m}{2} \right) \oao{a}{b} 
\times \mc^{j}_{m} \oao{a}{b} \right) 
\left( \bar{Ch}_c \left(s, -\frac{m+2\gamma}{2} \right) \oao{a}{b} 
\times \bar{\mc}^{j}_{-m+2\gamma} \oao{a}{b} \right) \right. 
\quad \nonumber \\
+ \ \left( Ch_c \left(s, -\frac{m}{2} \right) \oao{a}{b} 
\times \mc^{j}_{m+k} \oao{a}{b} \right) 
\left. \left( \bar{Ch}_c \left(s, -\frac{m+2\gamma}{2} \right) \oao{a}{b} 
\times \bar{\mc}^{j}_{-m+2\gamma+k} \oao{a}{b} \right) \right\}
\nonumber \\
\end{eqnarray}
Now we use the decomposition of extended continuous characters, 
eq.~(\ref{extmassiv}), and obtain:
\begin{eqnarray}
\frac{\log \epsilon}{4k\pi} \sum_{a,b \in \zi_2}
  \   \sum_{\gamma, \delta \in \zi_k} \sum_{2j=0}^{k-2}
\int_{0}^{\infty} ds \ 
\sum_{m \in \zi_{2k}} \nonumber \\
\times \ \left\{ \left(  \frac{q^{\frac{s^2}{k}}}{\eta^3} 
\Theta_{-m,k} \vartheta\oao{a}{b} 
\times \mc^{j}_{m} \oao{a}{b} \right) 
\left(  \frac{\bar{q}^{\frac{s^2}{k}}}{\bar \eta^3} 
\bar{\Theta}_{-m+2\gamma,k}
\bar{\vartheta} \oao{a}{b} 
\times \mc^{j}_{-m+2\gamma} \oao{a}{b} \right) \right. 
\quad \nonumber \\
+ \left. \left( 
\frac{q^{\frac{s^2}{k}}}{\eta^3} 
\Theta_{-m,k} \vartheta\oao{a}{b} 
\times \mc^{j}_{m+k} \oao{a}{b} \right) 
\left(  \frac{\bar{q}^{\frac{s^2}{k}}}{\bar \eta^3} 
\bar{\Theta}_{-m+2\gamma,k}
\bar{\vartheta} \oao{a}{b} 
\times \mc^{j}_{-m+2\gamma+k} \oao{a}{b}
\right) \right\}
\nonumber \\
\end{eqnarray} 
Using the decomposition of $SU(2)$ characters in terms of characters 
of the $N=2$ minimal models:
\begin{equation}
\chi^{j} (\tau) 
\ \vartheta \oao{a}{b} (\tau, \nu )
=  \sum_{m \in \zi_{2k}} 
\mc^{j}_{m} \oao{a}{b} (\tau, \nu )\ \Theta_{m,k} (\tau , -2\nu /k ), 
\end{equation}
and the charge conjugation symmetry of the $N=2$ minimal model characters,
 eq~(\ref{conjmin2}),  we get finally (see also~\cite{Petersen:1990qh}):
\begin{eqnarray}
\frac{\log \epsilon}{2\pi} \sum_{a,b \in \zi_2}
  \   \sum_{2j=0}^{k-2}
\int_{0}^{\infty} ds \ (q \bar{q})^{\frac{s^2}{k}}
\chi^{j} \bar{\chi}^j \vartheta^2 \oao{a}{b} 
\bar{\vartheta}^2 \oao{a}{b}
\end{eqnarray}  
Putting the extra factors, one obtain for the full DSLST 
background:
\begin{eqnarray}
\frac{\log \epsilon}{2\pi} \sum_{a,b \in \zi_2}
  \   \sum_{2j=0}^{k-2}
\int_{0}^{\infty} ds \ (q \bar{q})^{\frac{s^2}{k}}
\chi^{j} 
\bar{\chi}^{j}
\vartheta^2 \oao{a}{b} 
\bar{\vartheta}^2 \oao{a}{b}\ 
\frac{1}{(8\pi^2 \tau_2 )^2 \eta^4 \bar{\eta}^4}
\frac{\vartheta^2 \oao{a}{b}}{\eta^2} \frac{\bar \vartheta^2 \oao{a}{b}}{
\bar \eta^2}
\end{eqnarray}  
This is actually the type $0B$ partition function 
for the little string theory~\cite{Antoniadis:1994sr}, 
i.e. the coincident five-branes background. This 
result is intuitive. In fact, the spectra of the 
asymptotic states -continuous representations - 
match because the asymptotic geometry is insensitive 
to the fact that the five-branes are distributed on 
a circle in the transverse space.


\subsection{Spacetime supersymmetry}
In the previous section we have combined the $\slr / U(1)$ 
and the $SU(2) / U(1)$ supersymmetric coset with a 
$\zi_k$ orbifold, such that the $N=2$ charges are 
all integer, or, equivalently, such that the states of the theory 
belongs to representations of the small $N=4$ algebra.
A spacetime-supersymmetric theory is obtained 
from a type $0$ theory by twisting with the operator 
$(-)^{F_L}$ acting on the left worldsheet fermion number. 
Let us consider separately the continuous and 
discrete part of the spectrum.\footnote{In this analysis 
we don't consider the case of the new sectors of states since 
we don't have a good explanation for them -- however the 
twisting of the theory can be done in the same way.}
\subsubsection*{Continuous representations} 
Starting with continuous representations of $\slr / U(1)$, 
we get the following piece of the partition function 
after the twisting: 
\begin{eqnarray}
Z_{DLST}^{C} = \frac{1}{2} \sum_{a,b \in \zi_2} (-)^{a+b} 
\ \frac{1}{2} \sum_{\bar{a},\bar{b} \ \zi_2} 
(-)^{\bar a+ \bar b+ \eta \bar{a}\bar{b}} 
 \sum_{2j=0}^{k-2} 
\int_{0}^{\infty} ds \ \sum_{m,\bar{m} \in \zi_{2k}} 
\ \rho(s;m,\bar{m};a,\bar{a}) \nonumber \\
\times \ \left( Ch_c \left(s, -\frac{m}{2} \right) \oao{a}{b} 
\times \mc^{j}_{m} \oao{a}{b} \right) 
\left( \bar{Ch}_c \left(s, -\frac{\bar{m}}{2} \right) 
\oao{\bar a}{\bar b} \times 
\mc^{j}_{\bar{m}} 
\oao{\bar{a}}{\bar{b}} \right) 
\quad \nonumber \\
\nonumber \\
\times \ \frac{1}{(8\pi^2 \tau_2 )^2 \eta^4 \bar{\eta}^4}
\frac{\vartheta^2 \oao{a}{b}}{\eta^2} 
\frac{\bar \vartheta^2 \oao{\bar a}{\bar b}}{
\bar \eta^2}
\end{eqnarray}
In this expression, $\eta=0$ for type IIB superstrings 
and $\eta=1$ for type IIA superstrings.
As we have already noticed in the type 0B partition function, 
the leading divergent part of the density of states gives  
the partition function for the coincident NS5-brane background. 
\subsubsection*{Discrete representations} 
We now consider the piece coming from the discrete representations 
of $\slr / U(1)$. In the same way, we obtain the following 
expression:
\begin{eqnarray}
Z_{DLST}^{D} = \frac{1}{2} \sum_{a,b \in \zi_2} (-)^{a+b} 
\ \frac{1}{2} \sum_{\bar{a},\bar{b} \ \zi_2} 
(-)^{\bar a+ \bar b+ \eta \bar{a}\bar{b}} 
  \sum_{2j=0}^{k-2}
\sum_{2j'=1}^{k+1} \Upsilon (2j',1  ) 
\nonumber \\ \times \ \sum_{m,\bar{m} \in \zi_{2k}} 
\left( Ch_d \left(j', -\frac{m}{2} -j' \right) \oao{a}{b} 
\times \mc^{j}_{m} \oao{a}{b} \right) 
\left( \bar{Ch}_d \left(j', -\frac{\bar{m}}{2} -j' \right) 
\oao{\bar a}{\bar b} \times 
\mc^{j}_{\bar{m}} 
\oao{\bar{a}}{\bar{b}} \right) 
\quad \nonumber \\
\times \ \frac{1}{(8\pi^2 \tau_2 )^2 \eta^4 \bar{\eta}^4}
\frac{\vartheta^2 \oao{a}{b}}{\eta^2} 
\frac{\bar \vartheta^2 \oao{\bar a}{\bar b}}{
\bar \eta^2}
\nonumber \\
\end{eqnarray}
From this discrete spectrum we can compute the Witten index of 
the theory, that counts the number of Ramond ground states, 
for the $\slr / U(1) \times SU(2) / U(1)$ part of the background.  
We recall that the contribution to the Witten index of, 
respectively, the discrete extended 
characters of the supersymmetric $\slr / U(1)$ and the 
$N=2$ minimal model characters are: 
\begin{equation}
Ch_d (j',r) \oao{1}{1} (\tau,0 ) = \delta_{r+1,0 \ \mathrm{mod} \ k}
\ \ \ \mathrm{and} \ \ \  
\mc^{j}_{m} \oao{1}{1} (\tau , 0 ) = \delta_{2j,m}
\end{equation}
Therefore, the Witten index for the DSLST is: 
\begin{equation}
\sum_{2j=0}^{k-2} \sum_{2j'=1}^{k+1} 
\Upsilon (2j',1\ \mathrm{mod}\ k ) 
\sum_{m,\bar{m} \in \zi_{2k}} 
\delta_{2j,m} \ \delta_{2j,\bar{m}} 
\times \delta_{m-2j'+2,0} \ \delta_{\bar{m}-2j'+2,0} = k-1.
\end{equation}
Note that our result is in agreement with~\cite{Yamaguchi:2000dz}. 
It matches the cohomology of the $A_{k-1}$ ALE space, which is 
dual to this worldsheet CFT.

\section{The NS5-branes on a circle and the exact coset CFT}
In this section we would like to match the holographic description 
of the little string theory in a double scaling limit -- the 
orbifold of $\slr / U(1) \times SU(2) / U(1)$ discussed in the 
last section -- and the concrete supergravity description of the 
Higgs phase of little string theory, i.e. a distribution 
of NS5-branes in the transverse space. 

\subsection{The supergravity solution}
Following~\cite{Sfetsos:1998xd}, the supergravity for 
$k$ NS5 branes spread over a circle in the transverse space 
with $\mathbb{Z}_k$ symmetry is (in the string frame):\footnote{
We consider a special case of~\cite{Sfetsos:1998xd}, 
where $k$ stacks of N five-branes with a $\zi_k$ symmetry 
were considered. We restrict to the special case $N=1$.}  
\begin{equation}
ds^2 = \eta_{\mu \nu}dx^{\mu}dx^{\nu} 
+ H (\rho,\psi ) (dr^2 + r^2 d\phi^2 + d\rho^2 + \rho^2 d\psi^2 )
\end{equation}
where the conformal factor is given by the following sum with
$\mathbb{Z}_k$ symmetry:
\begin{eqnarray}
H &=& 1 + \sum_{n=0}^{k-1} \frac{\alpha' }{r^2 + 
\rho^2 + \rho_{0}^2 
-2\rho_0 \rho \cos (2\pi n/k - \psi)} \nonumber \\
&=& 1+ \frac{\alpha' k}{2\rho_0 \rho \sinh x} 
\Lambda_k (x,\psi )
\end{eqnarray}
with: 
\begin{equation}
e^x = \frac{r^2 + \rho^2 + \rho_{0}^2}{2\rho_{0} \rho} 
+ \sqrt{\left(\frac{r^2 + \rho^2 + \rho_{0}^2}{2\rho \rho_{0}}
\right)^2-1}
\end{equation}
and 
\begin{equation}\Lambda_{k} (x,\psi ) = 
\frac{\sinh kx}{\cosh kx - \cos k\psi }.
\label{instexp}
\end{equation}
This function captures the discreteness of the distribution of 
five-branes with $\zi_k$ symmetry. 
In the large $k$ limit, $\Lambda_k (x,\psi ) \to 1$, 
and we obtain a continuous 
distribution of NS5-branes.
We now take the double scaling limit:\footnote{Here $x^i$, $i=6,\cdots,9$ are
the Cartesian coordinates for the space transverse to the five-branes.}
\begin{equation}
g_s \to 0, \ \alpha' \ \mathrm{fixed}, \ 
U_0 \equiv \frac{\rho_0}{g_s \alpha'} \ \mathrm{fixed}, \ 
u_i \equiv \frac{x^i}{g_s \alpha'}\ \mathrm{fixed}
\end{equation}
Then, defining the coordinates: 
\begin{eqnarray}
u_6 &=& \rho_0 \sinh \rho \cos \theta \cos \varphi \ , \ \ 
u_7 = \rho_0 \sinh \rho \cos \theta \sin \varphi \nonumber \\
u_8 &=& \rho_0 \cosh \rho \sin \theta \cos \psi  \ , \ \ u_9 = \rho_0 \cosh
\rho \sin \theta \sin \psi \ ,
\end{eqnarray}
The solution is given by:
\begin{eqnarray}
ds^2 &=& \eta_{\mu \nu}dx^{\mu}dx^{\nu} + 
k \Lambda_k (x,\psi ) \left[ 
d\rho^2 + d\theta^2 + \frac{
\tan^2 \theta \ d\psi^2+ \tanh^2 \rho \ d\varphi^2
}{1+\tan^2 \theta \ \tanh^2 \rho} \right] \nonumber \\
B &=&
\frac{k \Lambda_k (x,\psi )}{1+\tan^2 \theta \ \tanh^2 \rho}\ 
d\varphi \wedge d\psi  + \frac{k \Lambda_k (x,\psi ) \sin k\psi}{
\tan \theta \, \sinh kx} \ d\varphi \wedge d \theta
\nonumber \\
e^{-2\phi} &=& \frac{\alpha'}{k}  U_{0}^2 
 \Lambda^{-1}_k (x,\psi ) \left( \cos^2 \theta 
\cosh^2 \rho + \sin^2 \theta \sinh^2 \rho \right)
\label{solcircl}
\end{eqnarray}
In the double scaling limit we have: $e^x = \cosh \rho / \sin \theta$.

\subsection{The coset description}
Our next task is to show that this supergravity background, 
eq.~(\ref{solcircl}), corresponds actually to an exact 
coset conformal field theory. In the following 
we will consider the supersymmetric coset 
construction in order to compare directly with the previous 
results. 
\subsubsection*{Superspace action for the coset}
We consider the $SL(2,\mathbb{R})_k \times SU(2)_k$ superconformal 
theory.
The supersymmetric WZW model is given by the following 
action for the matrix superfield $\mg$, taking values 
in $\slr \times SU(2)$:
\begin{equation}
kS (\mg ) = \frac{k}{2\pi} \int d^2 z\ 
d^2 \theta \ \mathrm{Tr} \ \left( 
 \mg^{-1} D \mg \mg^{-1} \bar{D} \mg 
+  \int dt \left\{ \mg \partial_t \mg^{-1} \left( D\mg\bar{D}\mg^{-1} +
\bar{D} \mg D \mg^{-1} \right) \right\}\right)
\end{equation}
and the superfield $\mg$ is parameterized in Euler angles 
for both group factors as follows:
\begin{equation}
\mg = \mathrm{diag} \left(
e^{\frac{i\sigma_2}{2} \Phi_L} e^{\frac{\sigma_1}{2} S}
e^{\frac{i\sigma_2}{2} \Phi_R} , 
e^{\frac{i\sigma_2}{2} \tilde{\Phi}_L} e^{i\frac{\sigma_1}{2} \Omega}
e^{\frac{i\sigma_2}{2} \tilde{\Phi}_R} \right),
\end{equation}
where the superfields expanded in superspace-coordinates read:
\begin{equation}
\Phi_{L,R} = \phi_{L,R} + \theta \chi_{L,R} + \bar \theta
\tilde{\chi}_{L,R} + \theta \bar \theta f_{L,R},
\end{equation}
and so on. We obtain the following action for 
$\slr \times SU(2)$ at level $k$:
\begin{eqnarray}
kS (\mg ) &=& \frac{k}{4\pi} \int d^2 z\ 
d^2 \theta \ \left\{ DS\bar D S - D\Phi_L \bar{D} \Phi_L 
- D\Phi_R \bar{D} \Phi_R - 2 \cosh S \ D\Phi_L \bar{D} \Phi_R
\right. \nonumber \\
&&\left. + D\Omega \bar D \Omega  + D\tilde \Phi_L \bar{D} \tilde \Phi_L 
+ D\tilde \Phi_R \bar{D} \tilde \Phi_R + 
2 \cos \Omega \ D \tilde \Phi_L \bar{D} \tilde \Phi_R \right\}
\end{eqnarray} 
We would like to gauge 
the following $U(1) \times U(1)$ null isometries of 
$\slr \times SU(2)$: 
\begin{equation}
(G,G') \to \left(e^{i\Lambda \sigma_2} G e^{i M \sigma_2} , 
e^{i\Lambda \sigma_2} G' e^{i M \sigma_2} \right)
\end{equation}
The isometry belongs to the elliptic subgroup of $\slr$.\footnote{
Another gauging involving the hyperbolic subgroup gives 
the Nappi-Witten cosmological solution~\cite{Nappi:1992kv}.}
Because the generator of the isometries in $\slr \times SU(2)$ are 
null, the left and right gaugings are independent (see e.g.~\cite{Klimcik:1994wp} 
for examples of null gaugings). Therefore the resulting target 
space will be four-dimensional. Note that in order to get a 
globally well-defined gauging, the gauge symmetry acts 
on $U(1)$ symmetries in $\slr$ and $SU(2)$ with the same periodicity. 
Therefore, one defines this coset with the {\it single 
cover} of $\slr$ rather than the universal cover which is 
usually considered.\footnote{Since the relative normalization 
of the action of the gauge fields in $\slr$ and $SU(2)$ is fixed 
by anomaly cancellation condition, the structure of the coset 
is rather rigid.}
The action of the gauged WZW model is given by adding the chiral 
projections of two $U(1)$ super-gauge fields, $\ma$ and $\bar \ma$:
\begin{eqnarray}
kS (\mg | \ma , \bar \ma ) = kS (\mg ) 
\qquad \qquad \qquad \qquad  
\qquad \qquad \qquad \qquad 
\qquad \qquad \qquad \qquad  
\qquad \qquad \qquad  
\nonumber \\
+ \frac{k}{2\pi} \int d^2 z  \ d^2 \theta  \left\{
\ma \left( \bar D \Phi_R + \cosh S \ \bar D \Phi_L \right) 
+ \bar \ma  \left( D \Phi_L + \cosh S \ D \Phi_R \right) 
-   \cosh S \ \ma \bar \ma \right. \nonumber \\
+ \left. \ma \left( \bar D \tilde \Phi_R + \cos \Omega \ \bar 
D \tilde \Phi_L \right) 
+ \bar \ma  \left( D \tilde \Phi_L + \cos \Omega \ D \tilde \Phi_R \right) 
+ \cos \Omega \ \ma \bar \ma  \right\} \nonumber \\
\end{eqnarray}
Under the super-gauge transformations the superfields transform as follows:
\begin{equation}
\begin{array}{ccccccc}
\Phi_{L} &\to& \Phi_{L} + \Lambda & , & 
\Phi_{R} &\to& \Phi_{R} + M \\ 
\tilde{\Phi}_{L} &\to& \tilde{\Phi}_{L} + \Lambda  & , & 
\tilde{\Phi}_{R} &\to& \tilde{\Phi}_{R} + M \\ 
\ma &\to& \ma + D \Lambda & , &
\bar{\ma} &\to& \bar{\ma} + \bar{D} M\\
\end{array}
\end{equation}
Therefore, we can make the super-gauge choice $\Phi_{L,R}=0$. 
We also define new coordinates: 
$\tilde{\Phi}_{L,R} = \Psi \pm \Phi$, $S=2R$ and 
$\Omega = 2 \Theta$. After integrating out the super-gauge fields, 
we find the following sigma-model:
\begin{eqnarray}
S = \frac{k}{\pi} 
\int d^2 z \ d^2 \theta \left\{ 
DR \bar D R + D \Theta \bar D \Theta + \frac{ 
\cos^2 \Theta \cosh^2 R \ D \Psi \bar D \Psi 
+ \sin^2 \Theta \sinh^2 R \ D \Phi \bar D \Phi
 }{\cosh^2 R - \cos^2 \Theta}
\right. \nonumber \\
+ \left. \frac{ \sin^2 \Theta \cosh^2 R \left( D \Psi 
\bar D \Phi 
- \bar D \Psi D \Phi \right)}{\cosh^2 R - \cos^2 \Theta}
\right\}\nonumber \\
\end{eqnarray}
This sigma-model corresponds actually to the solution 
describing five-branes on a circle, eq.~(\ref{solcircl}), 
in the limit $\Lambda_k \to 1$. 
A dilaton is generated by the change of the measure in 
the path integral, whose expression is given in the supergravity 
solution. Note that this sigma model receives no perturbative 
corrections, to all orders in $1/k$, because of 
$N=(4,4)$ superconformal symmetry~\cite{Alvarez-Gaume:1981hm}. Non-perturbative corrections 
(i.e. worldsheet instantons) can however appear. Indeed, it 
has been remarked in~\cite{Sfetsos:1998xd} that the function 
$\Lambda_k$ taking into account the discreteness of the 
distribution of five-branes, eq.~(\ref{instexp}), can actually be 
rewritten as the following infinite expansion:
\begin{equation}
\Lambda_k (x, \psi ) = 1 + \sum_{m \neq 0} 
e^{-k ( |m|x-im\psi )},
\end{equation}
which resembles an expansion in worldsheet instantons 
contributions. Therefore, as was done in~\cite{Tong:2002rq} for the 
T-dual of ALF space (using complementary methods), 
it should be possible to capture the exact 
supergravity metric by considering instantons of the gauged 
WZW theory.

The link between this model and the $\slr / U(1) \times 
SU(2) / U(1)$ orbifold can be seen as follows: 
let us perform a duality transformation along the $U(1)$ isometry generated 
by the superfield $\Psi$. Using the standard T-duality 
rules~\cite{Buscher:qj}, one finds the following dual action:
\begin{eqnarray}
S &=& \frac{k}{\pi}
\int d^2 z \ d^2 \theta \left\{ DR \bar D R + D \Theta \bar D \Theta +
\tanh^2 R \frac{D\Psi \bar D \Psi}{k^2} \right. \nonumber \\
&&\left. + \ \tan^2 \Theta \left( D\Phi  + \frac{D\Psi}{k} \right)
\left( \bar D \Phi + \frac{\bar D \Psi}{k} \right)
+ R^{(2)} \ln \left( \cosh^2 R \ \cos^2 \Theta  \right) 
\right\}\nonumber \\
\end{eqnarray}
If we redefine $\Psi /k = \Psi '$ and  
shift the superfield $\Phi$: $\Phi + \Psi/k = \Phi' $, we get exactly 
the superspace action for $\slr / U(1) \times SU(2) / U(1)$. 
We can have an indication of the presence of the $\zi_k$ orbifold for 
the following reason. The coordinate $\psi$ has a periodicity: 
$\psi \sim \psi + 2\pi n $, $n \in \zi$, 
inherited from the $SU(2)$ group manifold. 
After the field redefinitions, the new coordinates have the periodicity:
$\psi' \sim \psi' + 2\pi n/k$, and 
$\varphi' \sim \varphi' + 2\pi m + 2\pi n/k$. 
Therefore we have an $\zi_k$ orbifold whose action 
is given by the identifications: $(\psi',\varphi') \sim 
(\psi'+2\pi n/k,\varphi'+ 2\pi n/k)$. This is  the geometrical 
interpretation of the $\zi_k$ orbifold performed in sect.~\ref{sectlst} 
to achieve spacetime supersymmetry. To further argue that the theories are 
indeed the same, we will use below the exact coset description we 
found to show that the spectrum of the coset
$\{\slr \times SU(2)\}/(U(1)_L \times U(1)_R)$ and of the orbifold 
$\{\slr / U(1) \times SU(2) / U(1)\}/\zi_k$ are explicitely the same.

\subsubsection*{Path integral analysis}
We can use the standard techniques (see~\cite{Schnitzer:1988qj} 
and~\cite{Tseytlin:1993my}) 
to obtain the spectrum of the CFT described by this supersymmetric coset. 
One starts with the path integral for the supersymmetric coset: 
\begin{equation}
Z = \int [d \mg ] [d\ma ] [d \bar \ma ] 
e^{- kS (\mg | \ma , \bar \ma )} 
\end{equation}
Then one can parameterize the super-gauge fields by superfields 
taking values in the gauged $U(1)$ subgroup $H$:
\begin{equation}
\ma = \mh D \mh^{-1} \ , \ \ 
\bar \ma = \tilde \mh \bar D \tilde{\mh}^{-1}, 
\label{gaugepar}
\end{equation}
and using the supersymmetric generalization of the 
Polyakov-Wiegmann identity~\cite{Polyakov:tt}, one can rewrite the 
path integral in a product form:
\begin{equation}
Z = \int [d \mg ] [d\mh ] [d\tilde \mh ] [dB] [dC]
e^{- kS (\mh^{-1} \mg \tilde \mh )+ kS (\mh^{-1} \tilde{\mh}) - S_{SG}},  
\end{equation}
with a super-ghosts action coming from the Jacobian 
of the change of variables~(\ref{gaugepar}):
\begin{equation}
S_{SG} = \frac{1}{\pi} \int d^2 z \ d^2 \theta \ 
B \bar D C - \bar B D \bar C
\end{equation}
with $B = \beta + \theta b$, $C = c + \theta \gamma$, and the 
same for the antiholomorphic superfields. Then one change the variable 
$\mh^{-1} \mg \tilde \mh \to \mg$ and gauge-fix: $\mh = 1$. In the 
specific case of a null gauging we are considering, the super-WZW 
action for the null subgroup actually vanishes: $S (\tilde \mh ) = 0$. 
In short, the path integral for the model is just given by the sum 
of a $\slr \times SU(2)$ piece and a superghost action:
\begin{equation}
Z = \int [d \mg ] [dB] [dC]
e^{- kS (\mg )- S_{SG}},  
\end{equation}
The theory possesses two chiral BRST currents. The BRST charge for 
the left-moving symmetry is, in terms of components:
\begin{equation}
\mq_{BRST} = \oint \frac{dz}{2\pi} \ 
:c\ (\mj^3 + \mi^3 ): + :\gamma \ (\psi^3+\chi^3):
\label{BRSTsym}
\end{equation}
where $\mj^3$ and $\mi^3$ are respectively the total 
$\slr$ and $SU(2)$ currents at level $k$, and 
$\psi^3$, $\chi^3$ are the associated fermions.
We have a similar BRST charge for the right-movers.

\subsection{The spectrum of the coset and comparison with the orbifold}
The previous analysis shows that the spectrum of the coset 
is the spectrum of the $\slr \times SU(2)$ supersymmetric WZW model times 
the super-ghosts contribution, constrained by the BRST
symmetry~(\ref{BRSTsym}). For simplicity let us consider NS ground states. 
We recall that we have to consider the single cover of $\slr$. The weights of 
the primaries of the coset will be then:
\begin{eqnarray}
L^{cs}_0 &=& \frac{c_2}{k}+ \frac{j'(j'+1)}{k} 
- \frac{w_L}{2} \ m  + \frac{k}{4}w_{L}^2 \nonumber \\
\bar{L}^{cs}_0 &=& \frac{c_2}{k}+ \frac{j(j+1)}{k} 
- \frac{w_R}{2} \ \bar{m} + \frac{k}{4}w_{R}^2 \nonumber \\
\label{speccos}
\end{eqnarray}
where $c_2$ is the second Casimir of $\slr$ and 
$w_{L,R}$ the left and right spectral flows 
(see~\cite{Maldacena:2000hw} for more details).
In the BRST cohomology, 
the eigenvalues $m/2$ of $\mj^3$, the $\slr$ Cartan, are 
constrained to be related to $m' /2$, those of $\mi^3$, the 
$SU(2)$ Cartan:
\begin{eqnarray}
m = - m' \equiv j' - r' \ , \ r'=0,\ldots ,2j \nonumber \\
\bar m = - \bar m ' \equiv j' - \bar r ' \ , \ \bar 
r'=0,\ldots ,2j 
\end{eqnarray}

Now let us refresh our memory about the spectrum of the primaries 
for the orbifold $\{\slr / U(1) \times SU(2) / U(1) \} /\zi_k$ 
considered in the previous sections. We also restrict our attention 
to states that are in the ground state of the NS sector. Then we have 
the following conformal dimensions:
\begin{eqnarray}
L^{orb}_0 &=& \frac{c_2}{k}+ \frac{j'(j'+1)}{k} 
+ \frac{(m-\gamma-kw_+ )^2}{4k} - \frac{m'^2}{4k} \nonumber \\
\bar{L}^{orb}_0 &=& \frac{c_2}{k}+ \frac{j(j+1)}{k} 
+ \frac{(m+\gamma+kw_+ )^2}{k} - \frac{(-m'+2\gamma)^2}{4k}, \nonumber \\
\end{eqnarray} 
and due to the $\zi_k$ orbifold we have the constraint: 
$m+m'-\gamma = kp$. We do the change of variables: 
$-m'+2\gamma \to \bar{m}'$, 
and find the spectrum:
\begin{eqnarray}
L^{orb}_0 &=& \frac{c_2}{k}+ \frac{j'(j'+1)}{k} 
+ \frac{w_+ - p}{2} \ m' + \frac{k}{4}(w_+ - p )^2 \nonumber \\
\bar{L}^{orb}_0 &=& \frac{c_2}{k}+ \frac{j(j+1)}{k} 
+ \frac{w_+ +p}{2} \ \bar{m}'  + \frac{k}{4}(w_+ + p)^2 \nonumber \\
\label{specorb}
\end{eqnarray}
To match the two spectra of primaries, eqs.~(\ref{speccos}) 
and~(\ref{specorb}), one has to identify the left and right 
spectral flows as follows: 
\begin{equation}
w_L = w_+ -p \ , \ \ w_R = w_+ + p
\end{equation}
and the matching of the complete spectra follows from the matching 
of the dimensions of the primaries, since the structure of the 
characters is the same.  

To summarize, we have proven that the background corresponding 
to NS5-branes on a circle, eq.~(\ref{solcircl}) is given 
by the worldsheet orbifold CFT $\{\slr / U(1) \times SU(2) / U(1)\}/\zi_k$.
This was an assumption in~\cite{Giveon:1999px}, used 
to study little string theory in a double scaling limit. In this 
section we have given strong arguments for this equivalence, using 
an exact coset description of the background, 
and we explicitly matched the spectrum with 
the orbifold theory.

\section{About spacetime supersymmetry and geometry}

From this specific example we can draw important conclusions 
for the construction of supersymmetric string backgrounds 
in general. Suppose one would like to construct a superstring 
background that corresponds to the spacetime geometry of the 
two-dimensional Euclidean black hole times some flat coordinates:
\begin{equation}
ds^2 = k(d\rho^2 + \tanh^2 \rho \ d\varphi^2) 
+ \eta_{\mu \nu}dx^{\mu} dx^{\nu},
\end{equation}
described by an $\slr_k / U(1) \times \mathbb{R}^{d,1}$ SCFT. 
Then one chooses an internal superconformal field theory such that 
the total central charge is $c=15$. In our example, the ``internal''
CFT was simply $SU(2)_k / U(1)$, but one can choose more complicated 
examples as well. For the moment this theory cannot be spacetime 
supersymmeterized because the spectrum of charges of the $N=2$ 
R-current is not integer. To achieve this goal, one has to 
perform a projection on integral charges, which is realized 
in string theory as an orbifold, because of the requirement of 
modular invariance. As in the example we studied in detail in this 
paper, the order of the orbifold scales like $k$. Therefore 
the effective geometry in which the low-energy effective theory 
lives is deeply altered. Let us illustrate this idea on a 
simple ``toy model''. We consider simply a $U(1)
\times U(1)$ theory at level $k$.
The spectrum of the zero modes is given by:\footnote{ 
Only in this part, $\alpha'$ is reinstalled in the 
formulas instead of having been set to two.}
\begin{eqnarray*}
L_0 /\alpha'&=& \frac{(n-kw)^2}{4\alpha' k} 
+ \frac{(n'-kw')^2}{4\alpha' k} \nonumber \\
\bar{L}_0 / \alpha' &=& \frac{(n+kw)^2}{4\alpha' k} 
+\frac{(n'+kw')^2}{4\alpha' k} .
\end{eqnarray*}
The spectrum of the low energy effective action is obtained 
in the limit $\alpha' \to 0$, with $\alpha' k$ fixed. 
In this limit, the left and right contributions are 
equal and corresponds to states with mass squared 
$m^2 = (n^2+n'^2)/(\alpha' k)$. The winding
sectors are very massive and hence decouple from the low 
energy action.
Now let us consider the $\zi_k$ diagonal orbifold 
of this theory. We have {\it (i)} a projection 
on the states with $n-n' \equiv kp$ (with $p$ integer)
and {\it (ii)} new twisted sectors. 
The resulting spectrum is:
\begin{eqnarray*}
L_0 /\alpha' &=& \frac{[n-k(w+\gamma/k)]^2}{4\alpha' k} 
+ \frac{[n-k(w'+\gamma/k)+kp]^2}{4\alpha' k} \nonumber \\
\bar{L}_0 / \alpha'  &=& \frac{[n+k(w+\gamma/k)]^2}{4\alpha' k} 
+\frac{[n+k(w'+\gamma/k)+kp]^2}{4\alpha' k} .
\end{eqnarray*}
Then in the effective theory the contributions to the masses of the 
physical states will be:
\begin{eqnarray*}
m^2 &=& 2 \times \frac{(n-\gamma)^2}{4\alpha' k} \nonumber \\
\bar{m}^2 &=& 2 \times \frac{(n+\gamma)^2}{4\alpha' k}.
\end{eqnarray*}
Note in particular that the twisted sectors survive to the 
low energy limit, precisely because the order of the orbifold 
is $k$. 
This is qualitatively the same 
behavior as in the $ \slr / U(1) \times SU(2) / U(1)$ CFT
considered in this paper.

Some years ago, the following paradox was pointed out 
(see~\cite{Bakas:1994ba}~\cite{Bergshoeff:1994cb}~\cite{Bakas:1995hc}) : 
some backgrounds of superstring theory should be 
supersymmetric, because of worldsheet extended 
superconformal symmetry. However, the corresponding supergravity 
solution is manifestly {\it not} supersymmetric. One of the most obvious 
examples which was considered was in fact the $\slr / U(1) 
\times SU(2) / U(1)$ background. The string frame metric and dilaton for 
this solution are:
\begin{eqnarray}
ds^2 &=&  k(d\rho^2 + \tanh^2 \rho \ d\varphi^2) 
+ k(d\theta^2 + \tan^2 \theta \ d\psi^2) \nonumber \\ 
e^{-2\phi} &=& \cosh^2 \rho \cos^2 \theta.
\label{doublecos}
\end{eqnarray}
Let us consider for example the dilatino variation for 
type IIB supergravity in this background. We have:
\begin{equation}
\delta \lambda = 
\left[ \gamma^{\mu} \partial_{\mu} \phi \ \sigma_3 
-  \frac{1}{6} H_{\mu \nu \rho} \gamma^{\mu \nu \rho} 
\right] \oaop{\eta_1}{\eta_2} 
= \sqrt{k} 
\left[ -\Gamma^{\rho} \tanh \rho + \Gamma^{\theta} 
\tan \theta \right] \sigma_3 \oaop{\eta_1}{\eta_2}, 
\end{equation}
and therefore all the supersymmetry is explicitely 
broken. 

We now argue for a resolution of this 
paradox. The requirement of spacetime supersymmetry 
enforces the projection on integer charges in the 
underlying worldsheet CFT, hence the $\zi_k$ 
orbifold. Therefore the supergravity 
solution~(\ref{doublecos}) is {\it not} 
the correct low-energy effective background for 
this theory, since the spectrum we would 
get for supergravity fluctuations around the 
background~(\ref{doublecos}) wouldn't match
the $\alpha' \to 0$  limit  
(with $\alpha' k$ fixed)
of the $\left\{ \slr / U(1) \times SU(2) / U(1) \right\} 
/ \zi_k$ spectrum.
Rather the supergravity 
solution corresponding to this superstring 
background is given by the solution 
of five-branes on a circle, eq.~(\ref{solcircl}), 
which is manifestly supersymmetric.

\section{Conclusions}
In this work we have computed the torus amplitude 
for the $\slr_k / U(1)$ supersymmetric coset. We have used 
a supersymmetric marginal deformation method to obtain this result from the 
supersymmetric $\slr$ partition function, in analogy to~\cite{Giveon:1993ph}
for bosonic $SU(2)$ and~\cite{Israel:2003ry} for bosonic $\slr$. 
This is the first time this method is used for a supersymmetric WZW model. 
Then we have performed the exact decomposition in characters 
 -- and not only for the primary states -- of the supersymmetric coset, 
which are the same as the irreducible characters of the 
$N=2$ algebra with $c>3$. This last fact is an important element in 
favor of the duality between the $\slr / U(1)$ coset and the 
$N=2$ super-Liouville theory. Moreover we have seen that the 
{\it extended} characters of $N=2$, which have nice modular 
transformation properties, appear naturally in the decomposition.  
The spectrum we have found is composed of discrete representations, 
filling out exactly the improved range for the spin, and continuous representations 
accompanied by a divergent factor, due to the infinite volume available 
for them. Upon regularization, we find a divergent factor, 
corresponding to this volume divergence, and a finite part related 
to the reflection amplitude in $N=2$ super-Liouville theory.
Apart from these two pieces corresponding to standard characters 
of the coset, we find new sectors of states which may depend on the
regularization procedure. Note that this 
is not a particular problem of our computation, but would also 
appear in the bosonic coset partition function~\cite{Hanany:2002ev} ,
the $\slr$ partition function~\cite{Israel:2003ry} (and presumably
also in thermal $AdS_3$) if we 
also consider the {\it exact} expansion in characters, 
with the multiplicities, rather than only the expansion for the 
zero modes. This is an important open problem that deserves clarification.

Then we have considered the little string theory in a double 
scaling limit, corresponding to the worldsheet CFT
$\slr / U(1) \times SU(2) /U(1)$. The integrality 
of $N=2$ charges -- necessary to achieve spacetime supersymmetry -- 
is obtained by considering a $\zi_k$ orbifold of this CFT, much 
as in Gepner models~\cite{Gepner:1987qi}. We provide the 
spacetime supersymmetric partition function for this 
superstring background. The computation of the Witten index 
agrees with the expected result for superstrings on 
an $A_{k-1}$ ALE space. We also give some insight on the 
relationship between this theory and the theory of 
coincident NS5-branes, i.e. the $SU(2)_k \times \mathbb{R}_Q$ 
worldsheet CFT. We show in particular that the leading part 
of the partition functions, proportional to the (infinite) 
volume agree.

We have shown that the supergravity solution for NS5-branes 
spread on a circle in the transverse space with $\zi_k$ symmetry, 
given in~\cite{Sfetsos:1998xd}, is actually given by an exact 
CFT, corresponding to the null gauging: 
$(\slr \times SU(2))/ (U(1)_L \times U(1)_R)$. After 
integrating out the gauge field, we obtain a sigma-model 
which is perturbatively exact, thanks to $N=(4,4)$ worldsheet 
superconformal symmetry. The metric, dilaton and NS-NS two form 
background fields agree with the supergravity solution, up 
to a multiplicative factor taking into account the discreteness 
of the distribution of branes. This factor could be interpreted as 
an expansion in worldsheet instantons contributions, that correct the 
sigma-model of the coset. Using the standard BRST quantization 
of gauged WZW model, we have shown that actually the spectrum 
of the null coset coincides precisely with the spectrum 
of the $(\slr_{k} / U(1) \times SU(2)_{k} / U(1))/ \zi_{k}$ 
orbifold. This gives further strong arguments for the holographic 
duality between this worldsheet CFT and the double scaling 
limit of little string theory. 

Getting inspiration from this example, we have discussed the 
relationship between spacetime supersymmetry of superstring 
backgrounds and low-energy effective geometry. To achieve 
spacetime supersymmetry with the $N=2$ spectral flow one 
has to project onto integral charges of the $N=2$ R-current, 
and this projection has a drastic effect on the interpretation of the 
corresponding supergravity background. 
We have illustrated this idea with a very simple 
$U(1) \times U(1)$ example and of course with the 
$\slr_{k} / U(1) \times SU(2)_{k} / U(1)$ model. 
The correct supergravity description of the 
corresponding supersymmetric string background 
is the solution for NS5-branes on a topologically trivial circle. 
Also in the non-supersymmetric cases, it is not clear 
when it is possible to make sense of a string 
theory with non-integer $N=2$ charges since various 
OPE's between vertex operators of the theory will 
become non-local.

\section*{Acknowledgments}
Thanks to Costas Bachas, Ben Craps, Shmuel Elitzur, Amit Giveon, Elias Kiritsis, Anatoly Konechny, 
Marios Petropoulos, Boris Pioline, Sylvain Ribault, Volker Schomerus, Amit Sever, 
and Konstadinos Sfetsos for interesting discussions.
AP thanks the ENS-LPT for hospitality during the completion of this work.
AP is supported in part by the Horowitz Foundation. This work is 
supported in part by the Israeli Science Foundation.

\end{document}